\documentclass[
 aps,pra,
 amsmath,amssymb,
 reprint,
]{revtex4-2}

\usepackage{graphicx} 
\usepackage{dcolumn} 
\usepackage{bm} 
\usepackage{amssymb,amsmath,amsfonts,mathtools}
\usepackage{hyperref}

\usepackage[utf8]{inputenc}
\usepackage[T1]{fontenc}
\usepackage{mathptmx}
\usepackage{etoolbox}

\usepackage[english]{babel}
\usepackage[figuresleft]{rotating}

\usepackage{braket}
\usepackage{qcircuit}

\hypersetup{
	colorlinks = true,
	urlcolor   = [RGB]{0,172,172},
	linkcolor  = [RGB]{0,172,172},
	citecolor  = [RGB]{0,172,172}
}

\renewcommand*{\phi}{\varphi}
\renewcommand*{\epsilon}{\varepsilon}

\makeatletter
\def\@email#1#2{
 \endgroup
 \patchcmd{\titleblock@produce}
  {\frontmatter@RRAPformat}
  {\frontmatter@RRAPformat{\produce@RRAP{*#1\href{mailto:#2}{#2}}}\frontmatter@RRAPformat}
  {}{}
}
\makeatother

\begin{document}

\preprint{AIP/123-QED}

\title{Universal Broadband Linear Optical Transformations by an Interlaced Structured Integrated Photonic Processor}

\author{I.V.~Kondratyev$^{1}$}
\email{iv.kondratjev@physics.msu.ru}
\author{K.N.~Urusova$^{1}$}
\author{A.S.~Argenchiev$^{1}$}
\author{S.A.~Zhuravitskii$^{1}$}
\author{N.N.~Skryabin$^{1,2}$}
\author{A.D.~Golikov$^{3}$}
\author{V.V.~Kovalyuk$^{3,4}$}
\author{G.N.~Goltsman$^{2,3}$}
\author{I.V.~Dyakonov$^{1,2}$}
\author{S.S.~Straupe$^{1,2}$}
\author{S.P.~Kulik$^{1}$}

\affiliation{$^1$Quantum Technology Centre and Faculty of Physics, M.\,V. Lomonosov Moscow State University, 1 Leninskie Gory, Moscow, 119991, Russia}
\affiliation{$^2$Russian Quantum Center, 30 Bolshoy Boulevard, building 1, Moscow, 121205, Russia}
\affiliation{$^3$Department of Physics, Moscow State Pedagogical University, Moscow 119992, Russia}
\affiliation{$^4$Laboratory of Photonic Gas Sensors, University of Science and Technology MISIS, Moscow 119049, Russia}

\date{\today}

\begin{abstract}

Reconfigurable photonic integrated circuits enable applications in a wide range of fields. They represent a powerful platform for signal processing due to the ability to perform parallel computations. In this work, we demonstrate a photonic processor based on a six-channel universal chip  fabricated by femtosecond laser writing with an interlaced architecture comprising multiport beam splitters. We exploit and improve a previously demonstrated calibration procedure for a single building block \cite{kondratyev2025}, and extend it to a complete six-channel interferometer, reconstructing the corresponding interferometer models over a broad spectral range. We validate them by measuring a set of 100 Haar-random unitary matrices, achieving  mean amplitude fidelities of 95.9$\%$, 98.0$\%$, and 96.0$\%$ for radiation wavelengths 910, 945, and 980, respectively, without
on-device optimization. Furthermore, we experimentally realize spectral demultiplexing and multiplexing at these wavelengths, yielding routing fidelities ranging from 90.1$\%$ to 96.5$\%$ and total crosstalk values ranging from -9.6 to -14.4 dB, confirming the applicability of the fabricated photonic processor to practical wavelength-routing tasks.

\end{abstract}

\maketitle

\section{\label{sec:intro} Introduction}

Programmable multiport interferometers (PMIs) are compact, efficient and high-precision devices that enable the manipulation of optical field states \cite{bogaerts2020programmable}. They are characterized by low propagation losses, high stability, and high-speed optical operation, making them attractive for applications in communications \cite{Harris18, Zhou2024}. In addition, their energy efficiency, wide bandwidth, and ability to perform parallel operations make PMIs widely used in areas such as signal processing \cite{Perez2017, Xie24}, radio-frequency applications \cite{Zhuang15}, machine learning \cite{Shen2017, Zhang2021}, neuromorphic computing \cite{Tait17} and optical computing \cite{Yuhan23, Jiao2022}. Due to their scalability and ability to efficiently manipulate photons, PMIs also represent a promising platform for quantum simulators \cite{aspuruguzik2012, harris17, sparrow2018} and quantum computing \cite{PsiQuantum2025, Maring2024, Piergentili24}.

Programmable multiport interferometers are typically meshes of beamsplitters and phase shifters with $N$ input and output channels. The transformation applied by the interferometer to the input optical field is described by a $N \times N$ unitary matrix $U$, which can be controlled by setting the required phase shifts for the desired $U$. Of particular interest are universal multiport interferometers, capable of implementing any unitary transformation from the group $SU(N)$. This universality allows the same PMIs to be employed for a wide range of tasks \cite{Perez2017}. Widely-known examples of designs for universal multiport interferometers include Reck \cite{Reck1994} and Clements \cite{Clements16} schemes, both based on meshes of Mach–Zehnder interferometers - $2 \times 2$ interferometers composed of two balanced beam splitters and two phase shifters, which themselves can realize any unitary operation from $SU(2)$. A key advantage of these designs is the existence of analytical methods for determining the set of phases required to implement a desired unitary transformation, as well as algorithms for interferometer model reconstruction and phase shifter calibration. An interferometer model describes the dependence of the implemented transformation on the phase shifts $\vec{\phi}$, i.e. $U=U(\vec{\phi})$. Phase shifter calibration, in turn, refers to determining the relationship between the physical control parameter used to induce the phase shift and phase shift $\phi$. In the case of the thermo-optic phase shifters employed in this work, implemented as resistive heaters, the control parameter is the applied current $I$ (or, equivalently, the applied voltage), yielding the relationship $\phi=\phi(I)$ \cite{Ceccarelli19}. Together, they provide all the information required to implement arbitrary transformations.

Various platforms and fabrication techniques for PMIs have been demonstrated. For example, universal photonic chips based on $Si_3N_4$ fabricated using Triplex technology have been reported \cite{Taballione2021, Taballione2023, deGoede22}. Another possible material platform is silica-on-silicon, where a universal multiport chip was fabricated and successfully demonstrated as a photonic processor with high implementation accuracy, achieving an average amplitude fidelity of $99.9\%$ for $100$ Haar-random matrices \cite{Carolan2015}. However, femtosecond laser writing (FLW) is a widely adopted fabrication method due to its relatively simple, cost-effective, and rapid manufacturing process, as well as its unique ability to produce structures with three-dimensional geometries \cite{femtoCai, Skryabin2024}. Using FLW, a 24-mode optical chip was fabricated, demonstrating a mean amplitude fidelity of $99.7\%$ over $2000$ measured Haar-random matrices \cite{Barzaghi25}. In addition, 6-mode chips operating at wavelengths of $785$ nm and $1550$ nm with mean amplitude fidelities exceeding $99\%$ (for $1000$ Haar-random matrices) were demonstrated, showing potential for future applications in quantum computing and telecommunications \cite{Pentangelo24}. Recently, applications of photonic neural networks and high-dimensional boson samplers have been demonstrated using programmable optical chips manufactured using FLW technology, which feature a substantially three-dimensional waveguide structure \cite{cao2026programmable, di2026boson}.

\begin{figure*}[ht!]
\centering
\includegraphics[width=1\linewidth]{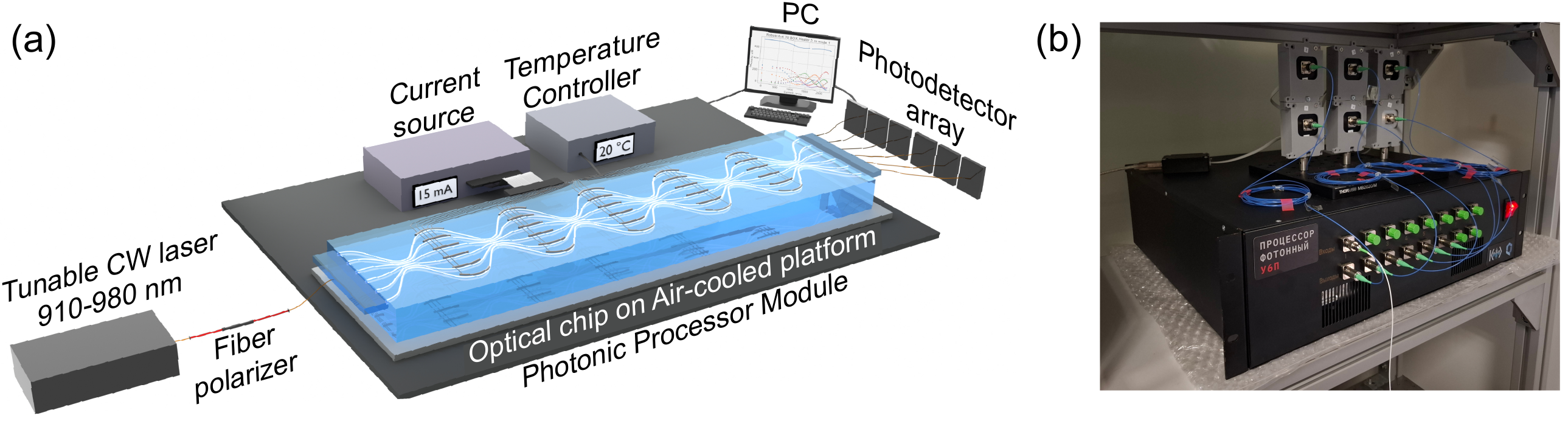}
\caption{(a) Schematic of the experimental setup. It consists of a tunable continuous-wave diode laser, photonic processor and six photodetectors. The processor contains a current source and an optical chip with a waveguide architecture based on \cite{Tanomura2020, kondratyev2020multiport}, mounted on a temperature-stabilized aluminum platform with a temperature setpoint of approximately 20$^\circ$C. The current source was connected to and fully controlled from a PC. (b) Photograph of the photonic processor and the six photodetectors connected to the output processor ports.} 
\label{fig:exp_scheme}
\end{figure*}

However, despite the advantages of the universal schemes of Reck \cite{Reck1994} and Clements \cite{Clements16}, their universality is highly dependent on the exact splitting ratios of the beam splitters. If the beam splitters are not perfectly balanced, these schemes lose this property \cite{kondratyev2020multiport}. As a result, PMIs are sensitive to imperfections arising during fabrication as well as to drifts occurring over long-term operation. In addition, since the splitting ratio of a beam splitter generally depends on the wavelength of the input optical field, these designs are universal only within a narrow spectral range. Several approaches aimed at mitigating this issue have been proposed \cite{Miller15, Burgwal17}. However, they typically lead to an increase in the complexity of the fabrication process as well as an increase in the total size of the scheme.

Therefore, an important challenge is the development of new universal designs that are robust against imperfections. To date, several such schemes have been proposed, in which the basic building blocks are not Mach–Zehnder interferometers but beam splitters \cite{Fldzhyan20} or multiport beam splitters \cite{Robust2020, Tanomura2020}. Due to their robustness to imperfections, these designs remain universal over a broad spectral range, which enables their use in applications involving multiple wavelengths, such as multiplexing and demultiplexing or parallel signal processing. Chips based on these designs have also been experimentally demonstrated \cite{tang2021ten, kondratyev2024large}.
However, the calibration procedure for the interlaced architecture \cite{kondratyev2020multiport, Tanomura2020} is considerably more challenging than that used for Reck and Clements designs, both in interferometer model reconstruction and phase shifter calibration. This is mainly because the transformation matrix of a $N\times N$ multiport beam splitter is, in general, parametrized by $N^2$ real parameters instead of one for the directional coupler. In addition, strong crosstalk between phase shifters complicates the calibration of the phase shift layers. Despite these challenges, methods for determining the matrices of the building blocks, namely multiport beam splitters, have been proposed \cite{kuzmin2021architecture, bantysh2023fast}, as well as an algorithm for the calibration of PMIs based on this design \cite{kondratyev2025}. However, the proposed calibration algorithm has so far been demonstrated only for a single building block of the architecture.

In this work, we present a 6-port photonic processor, a fully packaged PMI based on a universal complex scheme in a 19-inch case based on an integrated optical chip, which was fabricated using FLW technology. We demonstrate the successful use of the calibration algorithm proposed in \cite{kondratyev2025} at three different wavelengths, reconstruct the corresponding interferometer models, and validate them by measuring the amplitudes of 100 implemented Haar-random unitary transformations. As a result, average amplitude fidelities of 95.9$\pm$1.5$\%$, 98.0$\pm$0.6$\%$ and 96.0$\pm$2.3$\%$ were obtained for the three wavelengths, respectively, confirming the high quality of the reconstructed models. Importantly, all control parameters required to implement the target transformations were determined entirely from the reconstructed models, without performing any optimization directly on the photonic processor. Finally, we performed a (de)multiplexing experiment at the same three wavelengths. The results presented in this work demonstrate the potential of the proposed design \cite{Robust2020, Tanomura2020}, for the realization of universal programmable multiport interferometers and their practical applications.

\section{ Methods }
\subsection{Device fabrication} \label{subsec:fabric}

The optical chip of the photonic processor was fabricated using femtosecond laser writing of low-loss multiscan waveguides \cite{Skryabin2024} in a 100 mm $\times$ 50 mm $\times$ 5 mm fused silica glass sample (AGOptics, JGS1). To implement thermo-optical phase shifters (TOPS), microheaters were created above waveguides made of titanium (Ti) film deposited on the surface of the optical chip. More detailed information on the optical chip fabrication process is provided in Appendix~\ref{app:fabrication}.

Standard 8-channel PM fiber arrays with a pitch of 127 $\mu m$ were bonded to both sides of the optical chip using UV-curable optical adhesive (Norland NOA61). The optical chip was then placed on a thermally stabilized and air-cooled platform and housed in a 19-inch $\times$ 3U case (see Fig.~\ref{fig:exp_scheme}b), as were the temperature controller and electrical current controlling source for the TOPS. FC/APC fiber adapters (Diamond MPC-S8.22) were used as input/output portss for the processor, of which there were 8 pieces in reserve.

\subsection{Experimental setup} \label{subsec:setup}

The experimental setup is schematically shown in Fig.~\ref{fig:exp_scheme}a. Radiation from a tunable continuous-wave diode laser (Toptica CTL 950) is coupled into an input port of the processor by a polarization maintaining optical fiber patch-cord (PM780-HP) and a fiber polarizer \textcolor{black}{(930 nm In-line Polarizer)}, which ensures vertical polarization of the input radiation. The output powers at the chip output channels are measured using six photodetectors, which are connected to the output ports of the processor via fiber array.

\subsection{Interferometer model reconstructing and TOPS calibration}\label{subsec::calib}

\begin{figure}[ht!]
\centering
\includegraphics[width=1\linewidth]{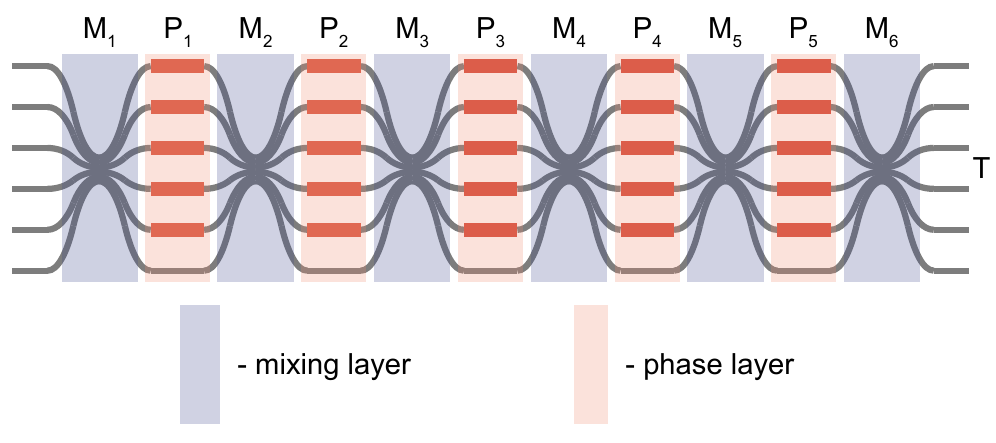}
\caption{Schematic of the investigated interferometer consisting of 6 multiport beam splitters and 25 phase shifters.} 
\label{fig:interferometer_scheme}
\end{figure}

The design of the interferometer is schematically shown in Fig.~\ref{fig:interferometer_scheme}. The interferometer consists of 6 multiport beam splitters, between which there are 5 phase layers with 5 phase shifters in each layer. The corresponding interferometer model is given by:
\begin{equation}
U = M_6 \prod_{i=1}^{5} (P_i M_i),
\label{eq:matrix}
\end{equation}
where $M_n,~n=\overline{1,6}$ are matrices of multiport beam splitters, and $P_m,~m=\overline{1,5}$ are matrices of phase layers containing the $m$-th vector of phase shifts between the modes, $\vec{\phi}_m=(\phi^{(m)}_1, \dots,\phi_5^{(m)}, 0)$: $P_m = diag(e^{i\phi_1}, \dots, e^{i\phi_5}, 1)$. We also account for the output losses. Since the model reconstruction is performed using normalized data (see Section~\ref{subsec::models}), only the output losses need to be included. To account for the output insertion losses, we introduce the diagonal matrix $T=diag(t_1,\dots t_6)$, where~$t_n$ represents loss of the $n$-th output channel. The complete interferometer model then becomes
\begin{equation}
U =T M_6 \prod_{i=1}^{5} (P_i M_i).
\label{eq:t_matrix}
\end{equation}

The phase shifts are introduced via the thermo-optic effect mentioned above. To induce a phase shift $\phi_i$ in the $i$-th mode, it is necessary to apply a current $x_i$ to the corresponding heater located directly above this channel. In the ideal case, the induced phase depends only on the current $x_i$, i.e., $\phi_i=\phi_i(x_i)$. However, in practice, crosstalk between TOPS within a single phase layer is present: when a current is applied, a heater induces a phase shift not only in the mode associated with the corresponding channel, but also in neighboring modes. In this case, $\phi_i$ becomes a function of all currents applied to the TOPS in this phase layer \cite{kondratyev2025}:
\begin{equation}
\phi_i = \phi_{0i} + \sum_j \alpha_{ij}\cdot x^2_j ,
\label{eq:phase}
\end{equation}

where each of the $\phi_{0i}$ represents the static phase shift in the particular interferometric arm, $x_j$ are the currents applied to the TOPS, and $\alpha_{ij}$ are the crosstalk coefficients describing the dependence between the current applied to the $j$-th heater and the phase shift introduced between the $i$-th mode and the reference mode.
Equation \ref{eq:phase} can also be rewritten in the matrix form:
\begin{equation}
\Phi = \Phi_0 + A\cdot X^2,
\label{eq:phases_mtx}
\end{equation}
where phase shifts and currents are expressed as column vectors $\Phi=(\phi_1, \dots,\phi_5)^T$, $\Phi_0=(\phi_{01}, \dots,\phi_{05})^T$, and $X^2 = (x_1^2, \dots, x_5^2)^T$, and crosstalk coefficients are collected in the matrix $A = {\alpha_{ij}}$.
This dependence holds for any phase layer $P_m$. Crosstalk is significant only between TOPS belonging to the same phase layer, i.e. within a single $P_m$. Crosstalk between different phase layers is negligible because the distance between adjacent phase layers (approximately 10 mm) is much larger than the spacing between neighboring waveguides within a phase layer (approximately 127 $\mu$m).

Thus, in order to fully control the transformation implemented by the interferometer, all the parameters introduced above must be determined. These include the matrices $T$ and $M_n$, $n=\overline{1,6}$, which define the interferometer model, and the calibration parameters $\Phi_{0m}$ and $A_m$, $m=\overline{1,5}$, which describe the dependence of the phase shifts on the applied currents. The interferometer model is used to determine, through optimization, the phase shifts required to implement a target transformation, while the calibration parameters provide the corresponding currents analytically. Since the interferometer model and the calibration parameters are reconstructed simultaneously and are intrinsically linked, we will hereafter refer to the complete set ${T, ~M_n,~\Phi_{0m},~A_m}$ simply as the model parameters.

It is also important to estimate the number of real parameters in our model. The matrix $T$ includes 6 parameters. Each matrix $M_n$ was parameterized using the well-known decomposition proposed in \cite{Reck1994}, resulting in 25 independent parameters in the range 0 to 2$\pi$. Thus, the complete set $\{M_n\}$ described by 25$\cdot$6=150 parameters, since each matrix is defined only up to arbitrary input and output phases. To determine the set of matrices ${P_m}$, it is necessary to know the initial phases for all layers (25 parameters in total, i.e., 5 phases for each of the 5 layers) as well as the crosstalk matrices (25$\cdot$5 = 125, since each $A_m$ has dimension 5$\times$5 and there are 5 layers). As a result, the six-mode interferometer is described by 6 + 150 + 25 + 125 = 306 real parameters.

\section{Experimental results}\label{sec::results}

\subsection{Characterization of the fabricated photonic processor} \label{app:static_char}

The fabricated photonic processor was characterized by measuring its insertion losses at different wavelengths, heater stability, and power consumption.

\begin{figure}[ht!]
\centering
\includegraphics[width=1.\linewidth]{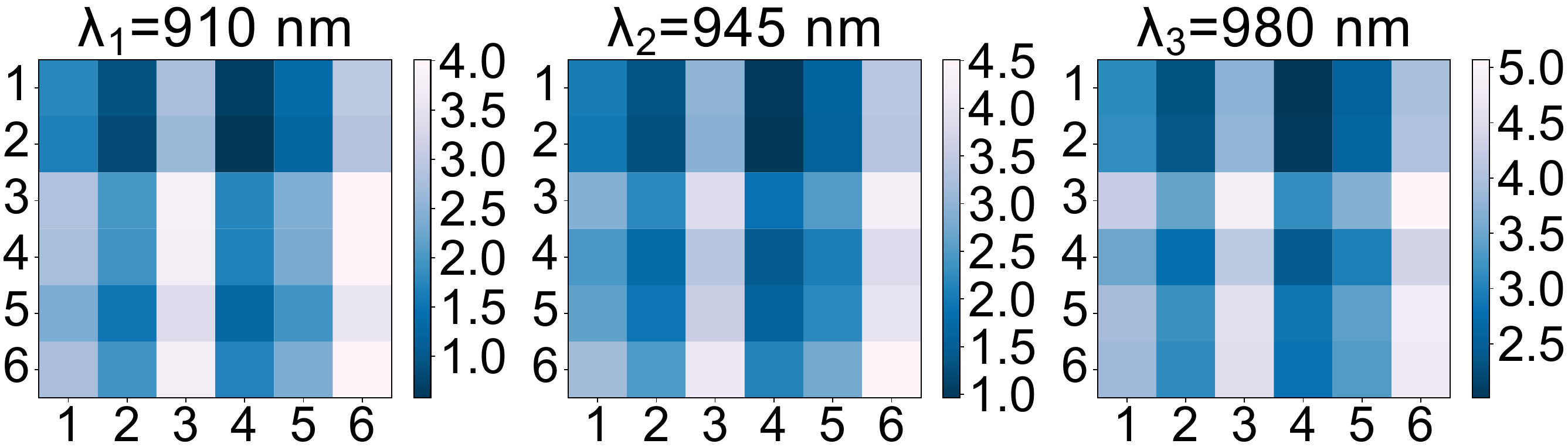}
\caption{Loss matrices measured at different wavelengths. The values are given in dB.}
\label{fig:loss_static}
\end{figure}

The measured insertion loss matrices are shown in Fig.~\ref{fig:loss_static}. The reported losses include both fiber-to-chip coupling losses and on-chip propagation losses. The average insertion losses are $2.3$, $2.5$ and $3.5$ dB at wavelengths of 910, 945 and 980 nm, respectively. The largest losses were observed at 980 nm because the waveguides were optimized for operation at 910 nm. As the operating wavelength moves away from the design wavelength, the propagation losses increase. For the same reason, the lowest insertion losses were obtained at 910 nm.

\begin{figure}[ht!]
\centering
\includegraphics[width=1.\linewidth]{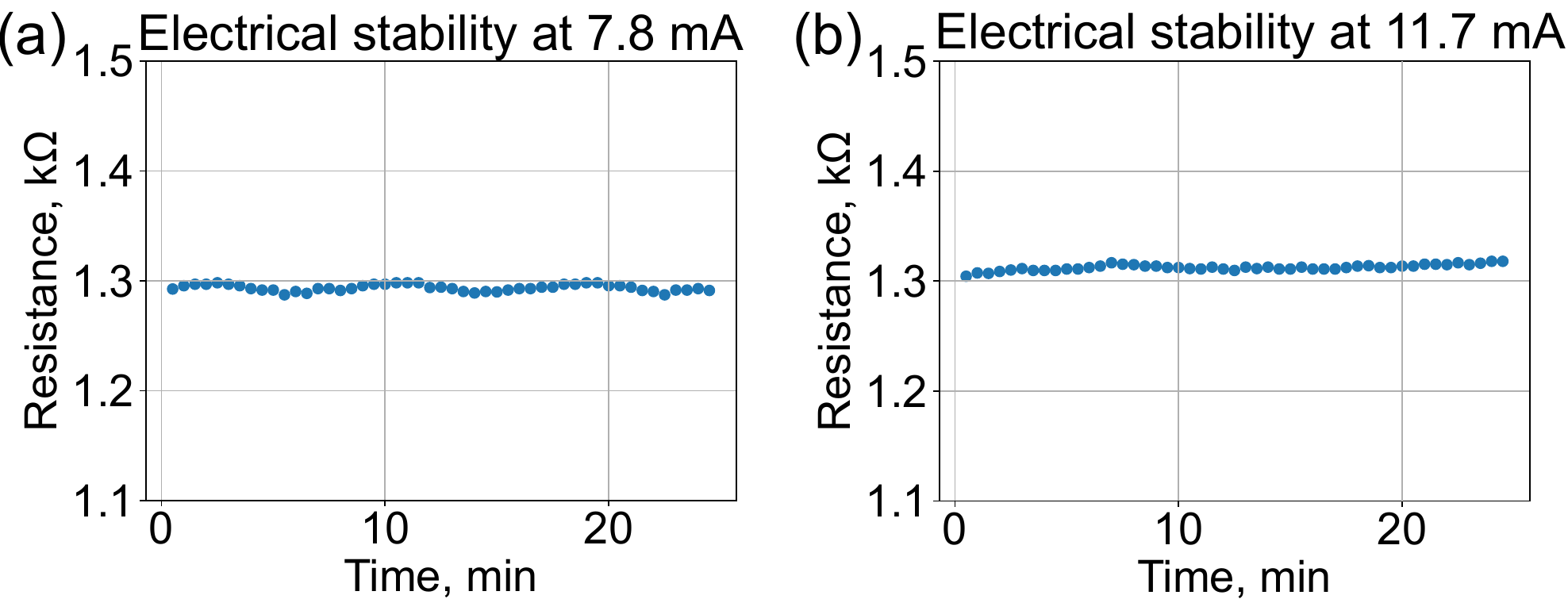}
\caption{Heater resistance stability measured over 25 minutes at (a) 7.8 mA and (b) 11.7 mA.}
\label{fig:res}
\end{figure}

We also investigated the temporal stability of the heater resistance in our fabricated PMI. For this purpose, a constant current of $7.8$ mA or $11.7$ mA (corresponding to dissipated powers of $79$ mW and $178$ mW, respectively) was applied to each individual heater, the voltage across this active heater was measured every 30 seconds and the corresponding resistance was calculated. This procedure was repeated for all heaters. The measured resistances varied by about 0.5$\%$ and 0.45$\%$ on average for the two current values, respectively, demonstrating the high electrical stability of the heaters. Moreover, the resistance exhibited small oscillations with no noticeable long-term drift, further confirming their stable operation (see Fig. \ref{fig:res})

Using the measured heater resistances, the total electrical power dissipation was estimated for all operating conditions considered in this work. The average total dissipated power was approximately 10 W.

Together, these measurements provide a comprehensive characterization of the fabricated photonic processor.

\subsection{Reconstructing the interferometer model}\label{subsec::models}

\begin{figure*}[ht!]
\centering
\includegraphics[width=1.\linewidth]{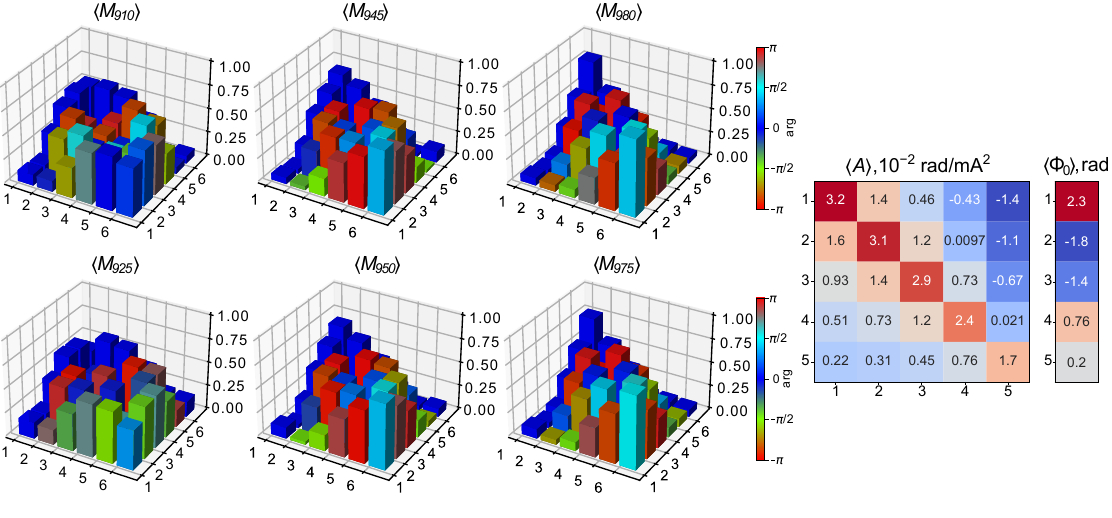}

\textcolor{black}{\caption{Model parameters for six wavelengths: $\lambda_1=910$ nm, $\lambda_2=945$ nm, $\lambda_3=980$ nm, $\lambda_4=925$ nm, $\lambda_5=950$ nm, and $\lambda_6=975$ nm. For each wavelength $\lambda_j$, the averaged multiport beam-splitter matrix $\langle M_{\lambda_j}\rangle$ was obtained by first aligning the matrices $M_n$ with respect to their input and output phases and then taking their arithmetic mean, followed by projection onto the nearest unitary matrix using SVD. The averaged initial phases $\langle\Phi_0\rangle$ and crosstalk matrix $\langle A\rangle$ were obtained by arithmetic averaging over all reconstructed $\Phi_{0m}$ and $A_m$, respectively, including all five phase layers and all six wavelengths.}}
\label{fig:models_M}
\end{figure*}

To reconstruct the interferometer model, specific experimental calibration data were first obtained. Coherent laser radiation was injected into the $i$-th input channel. The electrical currents $x_j$ in the range 0 to 22 mA  were then applied to the
$j$-th heater under calibration, and the output powers in all channels were measured. The size of an electrical current step was chosen to account for the quadratic dependence of the phase on the current in Eq.~\ref{eq:phase}, i.e., the current step decreased as the current increased. The measurements were performed on 43 current values within the range for each heater, which is an optimal number of data points for this protocol \cite{kondratyev2025}. This procedure was performed for all heaters and all input channels, resulting in a dataset ${i,~P(i,x_j, k),~x_j}$, where $P(i,x_j, k)$ is the power measured from the $k$-th output channel, $i=\overline{1,6}$, $j=\overline{1,25}$, $k=\overline{1,6}$. The resulting dataset was normalized by the total power measured at the chip output for each input channel (therefore, the values of 5 output channels were used). Consequently, only five output-channel powers are independent.

Since $P(i,x_j, k) \sim |U(x_j)|^2$, this dataset was used to reconstruct the model. The interferometer model was obtained by fitting the measured data using the model described in Section~\ref{subsec::calib}, where the interferometer transformation is given by Eq.~\ref{eq:t_matrix} and the current-to-phase dependence by Eq.~\ref{eq:phases_mtx}. The quality of the reconstruction was evaluated using the coefficient of determination ($R^2$), which characterizes the agreement between the measured data and the fitted model. Additional details are provided in Appendix \ref{app:fit}.

Alternatively, the calibration data can be measured more efficiently: by simultaneously sweeping five heaters from different phase-shifting layers. 
Since there is no thermal crosstalk between different phase-shifting layers for our photonic processor, this calibration approach would result in a five-fold speedup of the entire measurement time, as well as compactifying the experimental calibration data. This comes from the total number of $N/5$ calibrations, instead of $N$ calibrations for each input port of the chip, where $N$ is the total number of heaters under calibration. We provide more details on the simultaneous heater calibration in Appendix \ref{app:simultaneous_calibration_5X}. All measured calibration data together with the corresponding interferometer models are available at \cite{NN_Interferometer}.

After performing the calibration procedure described above, with a single heater being swept at a time, \textcolor{black}{models were obtained for three wavelengths}: 910 nm, 945 nm, and 980 nm, with coefficients of determination of 95.77$\%$, 96.50$\%$, and 97.67$\%$, respectively. The averaged values of $M_n$, $\Phi_{0m}$, and $A_m$ are shown in Fig. \ref{fig:models_M}. The matrices $M_n$ are identical up to input and output phase shifts. This is because the multiport beam splitters described by these matrices have identical geometry; therefore, their transformations are expected to be similar. However, deviations may arise due to fabrication imperfections.

\textcolor{black}{The matrices were compared as follows. The first matrix in each set of $M_n$ was chosen as the reference matrix. For each of the remaining five matrices, the input and output phases were optimized to maximize the fidelity between the reference and the considered matrix. The fidelity was calculated according to \cite{Pentangelo24}:
\begin{equation}
F_1 = \frac{1}{N}|Tr(U^{\dagger}V)|.
\label{eq:fid}
\end{equation}
This procedure was performed independently for all three sets of $M_n$. For the models at 910 nm, 945 nm, and 980 nm, the average fidelity with respect to the reference matrix} was 99.3$\%$, 99.5$\%$, and 99.7$\%$, respectively.

In addition, the obtained matrices $A_m$ are close to the estimates derived by analogy with the results reported in \cite{kondratyev2025} (see Appendix~\ref{app:alpha_estimation}). Together, these results confirm that the reconstructed models are physically consistent and correspond to the actual photonic processor. 

We have also performed the alternative calibration procedure, with five heaters being swept simultaneously at a time, and recovered three more models for the wavelengths 925 nm, 950 nm and 975 nm. The average coefficients of determination for these models are $96.5 \pm 2.8 \%$, $95.6 \pm 3.7 \%$ and $96.3 \pm 3.2 \%$, respectively. As we will discuss further in the text, these values could be noteceably increased by adding a term proportional to the fourth degree in the phase-current relationship (\ref{eq:phases_mtx}).

\begin{figure*}[ht!]
\centering
\includegraphics[width=1\linewidth]{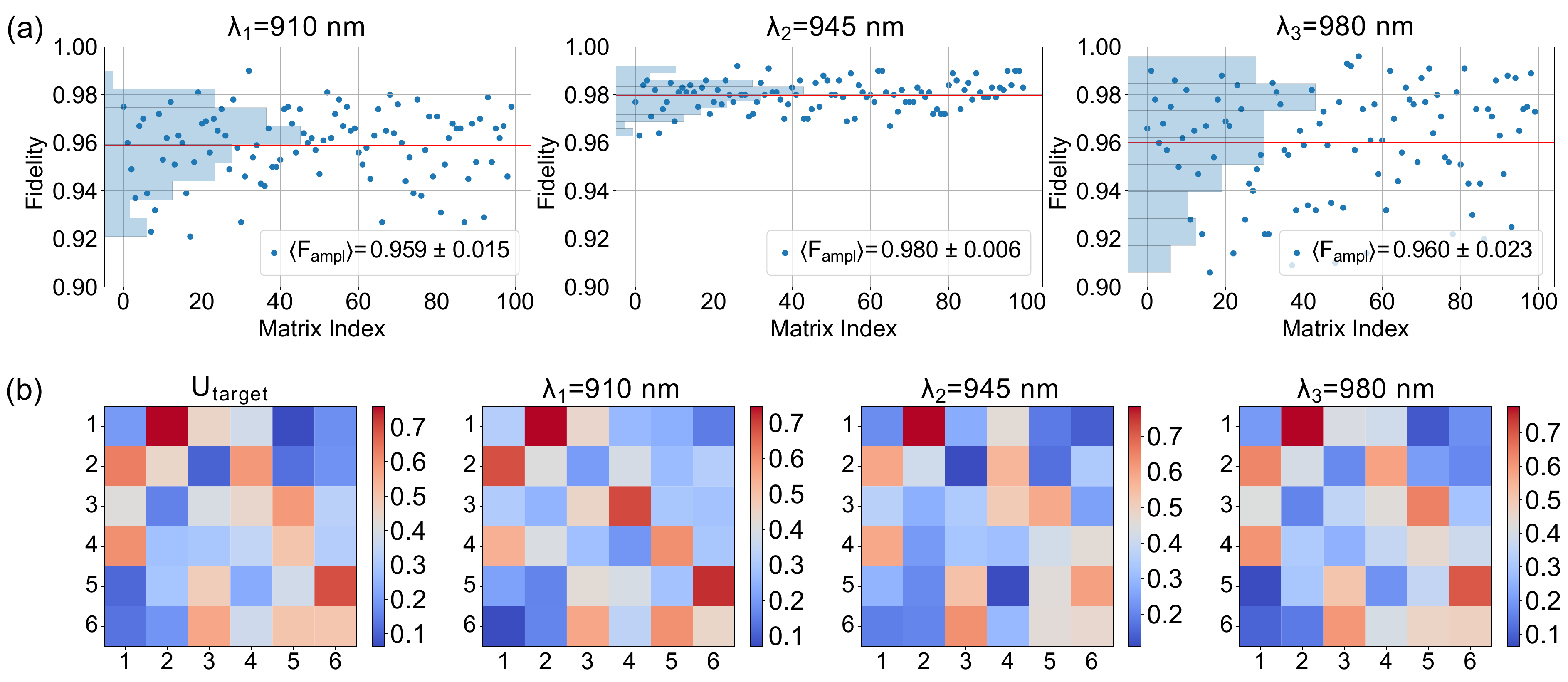}
\caption{(a) Distribution of the amplitude fidelity between $U_{meas}$ and $U_{target}$ for different wavelengths. The histogram represents the frequency of occurrence of the fidelity values, while the scatter plot shows their distribution over the indices of the measured matrices. The red line indicates the mean values of the distributions, which are 95.9$\pm$1.5$\%$, 98.0$\pm$0.6$\%$, 96.0$\pm$2.3$\%$ for 910 nm, 945 nm, and 980 nm, respectively.
(b) Examples of experimentally measured matrices for all three wavelengths. The fidelity between these matrices and $U_{target}$ is 96.1$\%$, 98$\%$, 99.6$\%$, respectively.} 
\label{fig:test}
\end{figure*}

\subsection{Testing the obtained interferometer models}\label{subsec:test}

After obtaining the models for all three wavelengths, it is necessary to verify that the reconstructed parameters accurately predict the behavior of the photonic processor under study. The validation was performed by implementing Haar-random unitary matrices and subsequently comparing the experimentally reconstructed amplitudes of the transformation matrix with the corresponding model predictions. The procedure was carried out as follows:

\begin{enumerate}
\item A Haar-random unitary matrix $U_{target}$ was generated according to \cite{HowHaar}.
\item \textcolor{black}{Using the obtained model, the phase shifts required to realize the target matrix $U_{target}$ up to input and output phase shifts were first found numerically. The resulting phase shifts were then converted into the corresponding heater currents using the calibrated current-to-phase relation. Additional details are provided in Appendix \ref{app:found}.}
\item The resulting set of currents was applied to the photonic processor.
\item Coherent radiation was injected into one of the input channels, and the output power distribution was measured.
\item The previous step was repeated for all input channels.
\item The measured output powers were normalized by the total output power for each input channel.
\item  The square roots of the normalized powers were taken to obtain the measured amplitude matrix $|U_{meas}|$. The resulting matrix was then compared with the target matrix $U_{target}$ using the amplitude fidelity \cite{Pentangelo24, Fyrillas2024}:
\begin{equation}
F_{ampl} = \frac{1}{N}\cdot Tr(|U_{target}^\dagger|\cdot|U_{meas}|),
\label{eq:fid_ampl}
\end{equation}
where $N$ is the number of channels; in our case, $N=6$.
\end{enumerate}

The procedure described above was carried out for all three wavelengths, using the same set of Haar-random matrices in each case. The dataset consisted of 100 matrices in total. As a result, 
distributions \textcolor{black}{of amplitude fidelity $F_{ampl}$} were obtained for all three models (see Fig.~\ref{fig:test}), with mean values of $95.9\pm1.5\%$, $98.0\pm0.6\%$, $96.0\pm2.3\%$, respectively. In addition, the model accurately reproduced the interferometer transformation in the zero-current configuration, yielding amplitude fidelities of $98.6\%$, $99.8\%$, and $98.9\%$ at 910 nm, 945 nm and 980 nm, respectively.

The results obtained show that the models presented in Section~\ref{subsec::models} are capable for predicting the transformation implemented by the universal multiport interferometer for a given set of currents. This implies that the interferometer matrix can be fully controlled at these particular wavelengths. 

\subsection{Demultiplexing experiment} \label{subsec:demulti}

\begin{figure}[ht!]
\centering
\includegraphics[width=0.9\linewidth]{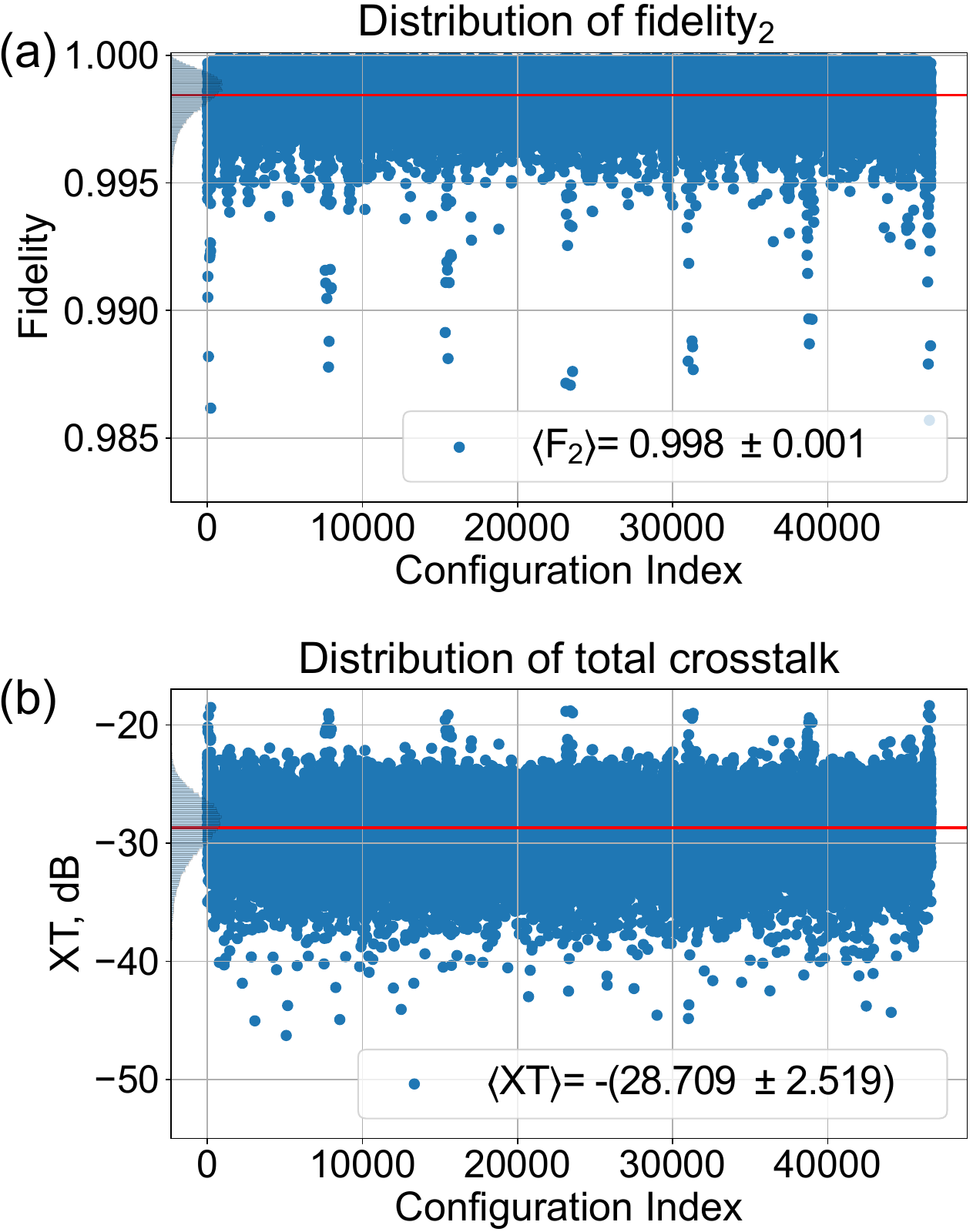}
\caption{(a) Distribution of the routing fidelity over all possible configurations of input and output channels. Mean value is $99.8\%\pm0.1$. (b) Distribution of the total crosstalk over the same set of routing configurations. Each point corresponds to one of the $6^6=46\ 656$ possible combinations of input and output channels for the three wavelengths. Mean value is $-28.7\pm2.5$ dB.}
\label{fig:rout}
\end{figure}

Having reconstructed and validated the interferometer models, we next demonstrated one of the potential applications of the fabricated photonic processor, namely spectral demultiplexing and multiplexing.

In the present context, spectral demultiplexing refers to routing different spectral components to different output channels, whereas multiplexing corresponds to combining them into a common output channel. Owing to the broadband universality of the employed interferometer design, the photonic processor can potentially implement not only these two particular operations but also a general wavelength-routing task, in which the input and output channels can be chosen independently for each spectral component. However, universality at each wavelength separately does not by itself guarantee that the desired transformations at several wavelengths can be realized simultaneously using a single current configuration. Therefore, this capability requires an additional verification.

For this purpose, a numerical routing experiment was performed using the reconstructed interferometer models at the three wavelengths. For each wavelength, an input channel and a desired output channel were specified, and a single current configuration common to all three models was numerically determined to realize the corresponding routing operation. This procedure was repeated for all possible combinations of input and output channels. Since both the input and output channels can independently take one of six values for each of the three wavelengths, the total number of routing configurations is $6^6=46 \ 656$. Details of the current-configuration optimization are provided in the Appendix \ref{app:rout}.

The routing quality was evaluated using two metrics. The first metric was the routing fidelity
\begin{equation}
F_2(X,Y)=\left(\sum_j\sqrt{X_j\cdot Y_j}\right)^2,
\label{eq:routing_fidelity}
\end{equation}
where $X$ and $Y$ are the target and obtained output power distributions, respectively. The second metric was the total crosstalk
\begin{equation}
XT=10\cdot\log_{10}\frac{\sum_i P_i}{P_{\mathrm{desired}}},
\label{eq:total_crosstalk}
\end{equation}
where $P_i$ denotes the power in the undesired output channels and $P_{\mathrm{desired}}$ is the power in the target output channel. Total crosstalk was included to facilitate comparison with devices specifically designed for wavelength-routing applications.

For each routing configuration, the output power distributions obtained at the three wavelengths were concatenated into a single vector together with the corresponding target distributions. The resulting vectors were normalized before calculating the routing fidelity in Eq.~\ref{eq:routing_fidelity}. The total crosstalk was evaluated in an analogous manner by combining the desired and undesired powers over all three wavelengths. The distributions of both metrics over all $46\ 656$ routing configurations are shown in Fig.~\ref{fig:rout}. The mean routing fidelity is $99.8\%$, while the mean total crosstalk is $-28.7$ dB. These results indicate that, according to the reconstructed models, the photonic processor can realize a wide range of wavelength-routing configurations with high fidelity and low crosstalk.

\begin{figure*}[ht!]
\centering
\includegraphics[width=0.9\linewidth]{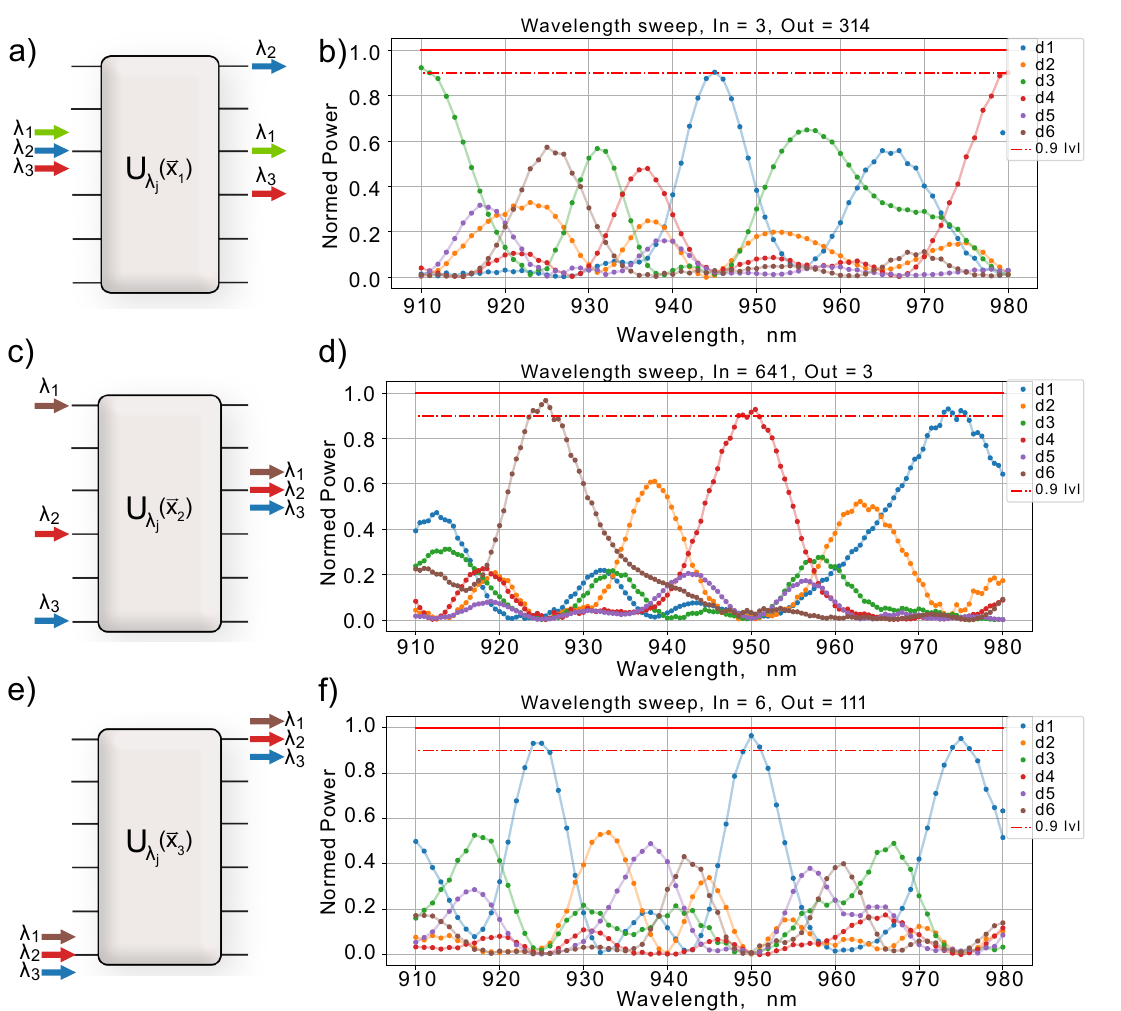}
\caption{(a, b) Scheme of the demultiplexing task and experimentally measured output power distributions for radiation injected into input channel 3. The wavelengths 910 nm, 945 nm, and 980 nm are routed to output channels 3, 1, and 4, respectively. (c, d) Scheme of the multiplexing task and experimentally measured output power distributions for radiation at 925 nm, 950 nm, and 975 nm injected into input channels 6, 4, and 1, respectively. All three wavelengths are routed to output channel 3. The wavelength sweep was measured using the reversed input/output ports. (e, f) Scheme of routing task and experimentally measured output power distributions for radiation injected into input channel 1. The wavelengths 925 nm, 950 nm, and 975 nm are all simultaneously routed to output channel 1. This particular routing task is one of the most challenging for the six-port Clements universal interferometer, due to the inevitable dependence of the DC splitting ratio on the wavelength. }
\label{fig:multi}
\end{figure*}

In addition to the numerical analysis of the general wavelength-routing problem, spectral demultiplexing and multiplexing were experimentally demonstrated.
For demultiplexing, radiation was injected into channel $3$ and routed to channels 3, 1 and 4 at wavelengths of 910 nm, 945 nm and 980 nm, respectively. For multiplexing, channels 6, 4 and 1 were used as inputs for the three wavelengths: 925 nm, 950 nm and 975 nm, respectively, and channel 3 was used as the common output.
We also performed a routing task for which all three wavelengths were simultaneously redirected from the sixth input port to the first output port. It should be noted that the latter routing task is one of the most challenging for the six-port Clements universal interferometer, due to the inevitable dependence of the directional coupler splitting ratio on the wavelength.
The corresponding current configurations for all experimental routing demonstrations were applied to the photonic processor, and the transformation matrices were measured using a procedure similar to that described in Section~\ref{subsec:test}. The results of the experimental demultiplexing and multiplexing are shown in Fig.~\ref{fig:multi} in the form of the experimental wavelength sweeps. 

As a result, the routing fidelities for demultiplexing are $92.3\%$, $90.3\%$, and $90.1\%$ for the three wavelengths, respectively, while for multiplexing they are $94.8\%$, $91.5\%$, and $92.2\%$. The resulting total crosstalk values were $-10.8$ dB, $-9.6$ dB, and $-9.7$ dB for demultiplexing, and $-12.6$ dB, $-10.3$ dB, and $-10.7$ dB for the multiplexing task. The routing fidelities of "all from 6 input to 1 output" are $93.1\%$, $96.5\%$, and $95.2\%$, and the corresponding crosstalk values are $-11.3$ dB, $-13$ dB, and $-14.4$ dB.

Although the crosstalk values obtained were not as favorable as those reported for dedicated wavelength-routing devices, functionality was demonstrated on a universal chip rather than on a device optimized for a single task. Overall, these results demonstrate the applicability of the fabricated photonic processor to multiwavelength routing and highlight one of its potential practical applications.

\begin{figure*}[ht!]
\centering
\includegraphics[width=1\linewidth]{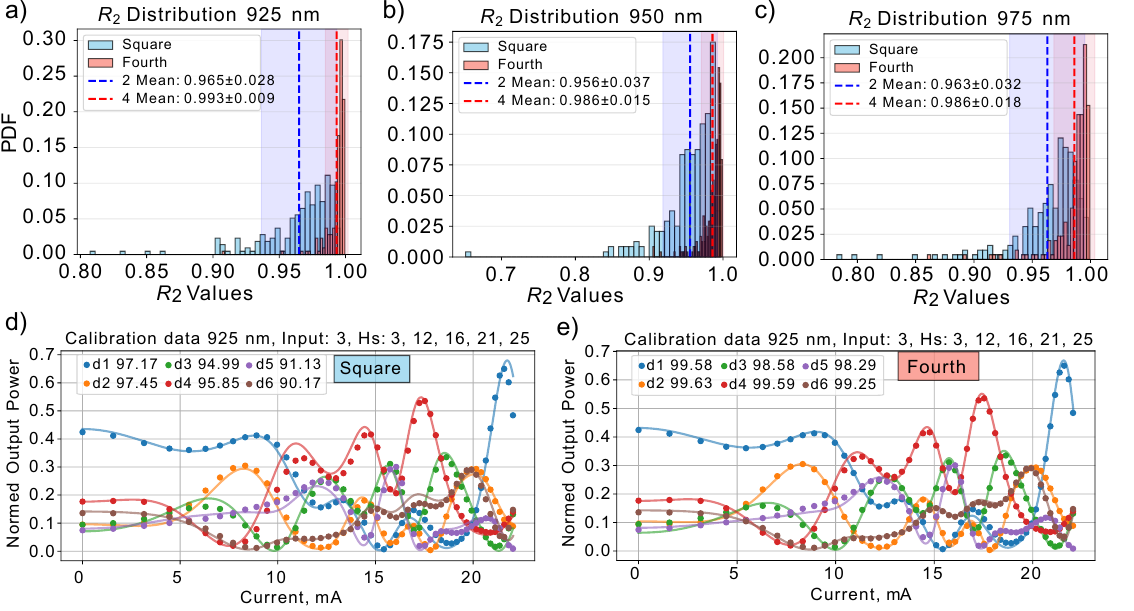}
\caption{ The distribution of the coefficients of determination $R^2$ for each individual experimental curve for the three wavelengths of the input radiation to the optical chip. The blue histogram corresponds to a model with a quadratic dependence of phase on current (\ref{eq:phases_mtx}), while the red histogram corresponds to a model with a fourth degree dependence of phase on current (\ref{eq:phases_mtx_fourth}). An example of a fit of experimental calibration data at a wavelength of 925 nanometers. d) The fit of experimental data with a model with a quadratic dependence of phase on current (\ref{eq:phases_mtx}), and e) shows the fit of experimental data with a model with a fourth degree dependence of phase on current (\ref{eq:phases_mtx_fourth}). The measured intensities normalized by the sum at the output of the optical chip are represented by dots, while the values predicted by the model are represented by continuous lines of the same color. The legend shows the values of the coefficients of determination for each individual experimental curve. The histograms of the distribution of these values are shown in a), b), c). This example clearly shows an improvement in the quality of the experimental data approximation for a model with a fourth degree of phase-current dependence. } 
\label{fig:R_2_Distributions}
\end{figure*}

\section{Discussion and conclusion}

We demonstrated a fully packaged photonic processor. Its use is not restricted to a specific optical setup or a particular experiment. To perform measurements, it is sufficient to connect the device to a source and measurement equipment; once the photonic processor has been calibrated and its interferometer model reconstructed for a given optical source, no additional optical alignment is required. This makes the photonic processor a stable and convenient device that can be employed in a wide range of experiments without assembling a dedicated setup. A similar packaged photonic processor was also demonstrated in \cite{Taballione2021, cheng2024multimodal}.

The previous use of the calibration algorithm considered was demonstrated only on a single building block of the design employed \cite{kondratyev2025}. In this work, we have successfully extended its use to a complete six-channel interferometer. Unlike approaches reported in \cite{kondratyev2024large, friedman2025programmable}, no on-device optimization was required for any of our experiments to implement target unitary transformations. Instead, all optimization was performed numerically using reconstructed interferometer models. Such an approach will extend device lifetimes and reduce resources required during operation.

In this work, we experimentally demonstrated the possibility of using a manufactured optical chip for simultaneous spectral demultiplexing of optical signals at three wavelengths. These functions are relevant for a variety of applications, including optical networks \cite{Keiser1999}, optic-based sensors \cite{Costley2016}, optical computing \cite{Xu2023, GironCastro2024},  parallel machine learning \cite{Aadhi2025}, and parallel signal processing \cite{Hong2025}. This further illustrates the practical utility of broadband universal photonic processors. 

Although a more detailed study of the spectral capabilities of the interferometer architecture in question, i.e., the interlaced structure, is outside the scope of this work, we would like to draw attention to some open questions. For example, the question may naturally arise whether optical signals can be spectrally separated for more than three wavelengths on such a photonic processor. On the other hand, using a chip for spectral separation of three wavelengths, it is reasonable to ask: how spectrally far apart can the wavelengths of the target signals be, and conversely, how close can the wavelengths of the target signals be located? A partial answer to the latter question can be suggested using experimentally measured wavelength sweeps shown in Fig. \ref{fig:multi}, as it can be seen from these graphs that the average characteristic peak width (from maximum to zero) is around 15 nanometers. This gives us a physical insight that the minimum allowable wavelength resolution for our device is 15 nm. Nevertheless, these and other questions show that the study of optical interferometers with an interlaced structure requires ongoing research, both experimental and theoretical.

A significant discrepancy between the experimentally measured fidelity values of the multi wavelength optical routing and the theoretically predicted ones, as well as the imperfect reconstruction of 100 random Haar matrices, may be due to the incompleteness of the digital model of the optical chip under study. For example, we observe a noticeable effect on the resulting quality of calibration data fitting from the addition of a term proportional to the fourth power of the currents applied to the phase current dependence (\ref{eq:phases_mtx}):
\begin{equation}
\Phi = \Phi_0 + A\cdot X^2 + B\cdot X^4.
\label{eq:phases_mtx_fourth}
\end{equation}
This effect is demonstrated in Fig. \ref{fig:R_2_Distributions}(a-c) where three histograms of the individual $R^2$ coefficient of each experimental curve are shown for three wavelengths. An example of the experimental data fitted by quadratic (\ref{eq:phases_mtx}) and fourth-degree (\ref{eq:phases_mtx_fourth}) phase-current models is shown in Fig d) and e). Adding a $ B\cdot X^4$ term in equation (\ref{eq:phases_mtx_fourth}) clearly improves the overall quality of the global curve fitting and potentially improves the experimental models of the optical chip. More details on the chip model improvements are given in Appendix~\ref{app:fouth}.

One possible experimental improvement to the fabricated photonic processor is the introduction of thermal insulation trenches \cite{Ceccarelli2020_low, Albiero2022}. Such trenches can reduce thermal crosstalk, resulting in significantly smaller off-diagonal elements of the crosstalk matrices. Although this would not eliminate the need to account for crosstalk in the interferometer model, it could enable operation at lower currents and therefore reduce the thermal load on the photonic processor. Another promising direction is the use of slightly more complex femtosecond laser writing techniques \cite{Pentangelo2022FLW} that enable fabrication of circuits with lower losses and higher refractive-index contrast, which could allow more compact optical chips. These approaches provide clear directions for the future development of the proposed photonic processor.

Finally, we believe that the results obtained in this work indicate that the design proposed in \cite{tang2017integrated, kondratyev2020multiport, zelaya2024goldilocks}, despite its increased complexity in model reconstruction and calibration compared to Reck \cite{Reck1994} and Clements \cite{Clements16} schemes, equipped with the complete programming algorithm proposed here, is a promising platform for the realization of broadband universal programmable multiport interferometers.

\begin{acknowledgments}           
The authors acknowledge support from Russian Science Foundation grant 22-12-00353-P (https://rscf.ru/en/project/22-12-00353/); V.K., A.G., and G.G. acknowledge support of the Ministry of Science and Higher Education of the Russian Federation FSME-2025-0004 (PICs fabrication).  

\end{acknowledgments}

\bibliography{PLib}

\appendix

\section{Optical chip fabrication details} \label{app:fabrication}

\begin{figure*}[ht!]
\centering
\includegraphics[width=0.9\linewidth]{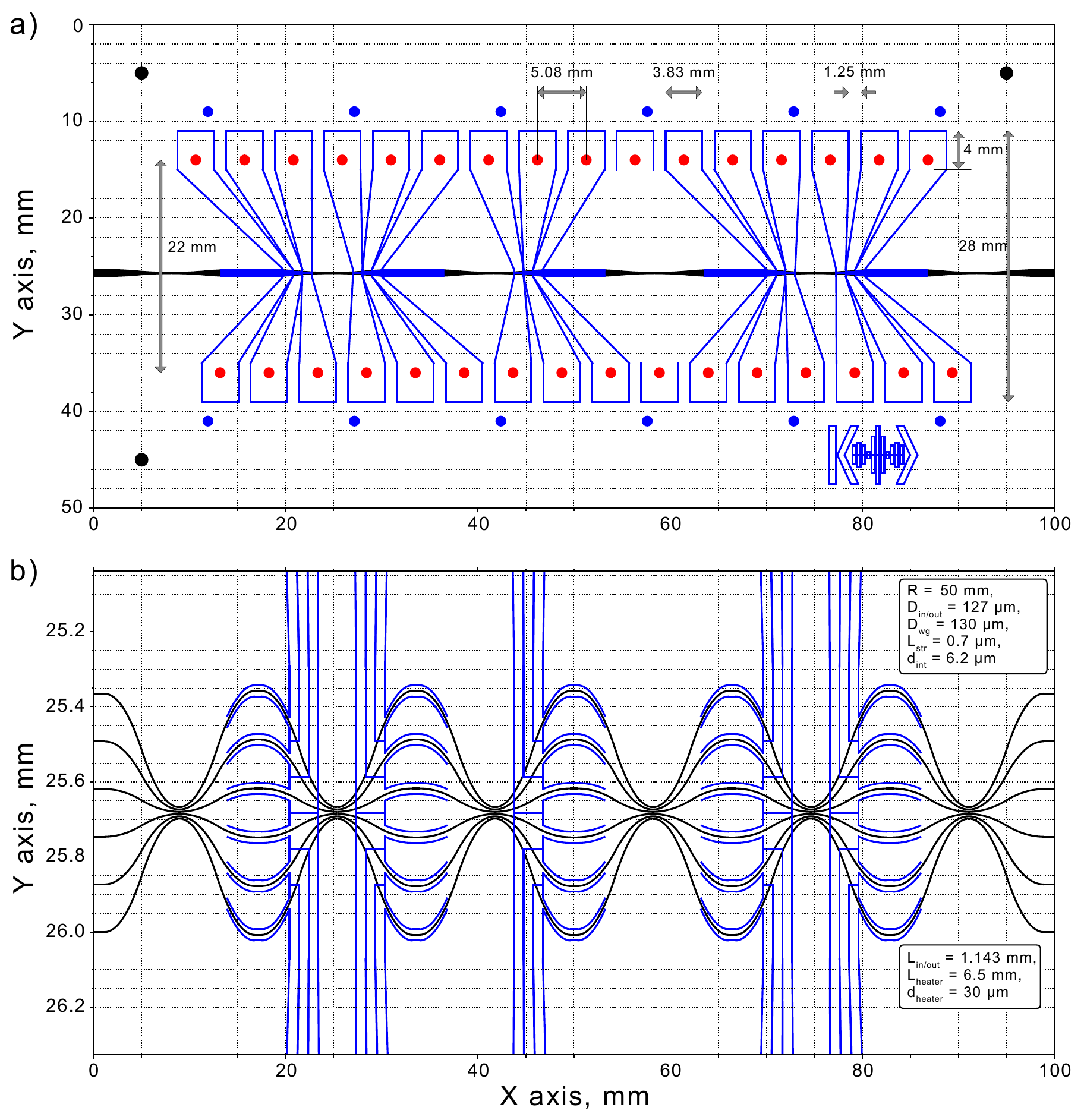}
\caption{Scheme of the optical chip. The black lines show the waveguides, and the black dots show the alignment marks used to align the waveguides and heaters. The blue lines show the pattern for engraving the insulating tracks on the deposited titanium film on the chip surface to form the contact pads, supply electrodes, and heaters. The red dots indicate the locations of the signal pogo-pins contact, and the blue dots for the ground pogo-pins.}
\label{fig:fabrication}
\end{figure*}

The optical chip of the photonic processor was fabricated using femtosecond laser writing of low-loss multiscan waveguides in a 100 mm $\times$ 50 mm $\times$ 5 mm fused silica glass sample. The optical chip scheme is shown in Fig.~\ref{fig:fabrication}. The waveguides (black lines) were written using second-harmonic femtosecond pulses of an ytterbium-doped fiber laser at a wavelength of 515 nm, a duration of 270 fs, an energy of 47 nJ at a frequency of 1 MHz. The pulses were focused using an aspheric lens with a numerical aperture of NA = 0.55 at a depth of 20~$\mu m$ below the surface of the sample. Each waveguide consists of $N = 21$ scans at a speed of 4 mm/s with a small offset of $s = 0.2 \mu m$ between scans.

The distance between adjacent input and output channels was $D_{in/out}$ = 127 $\mu m$ to match standard fiber arrays, while the distance between waveguides inside the interferometer was slightly increased to $D_{wg}$ = 130 $\mu m$. The multiport beam splitters were implemented as an array of coupled waveguides. The minimum radius of curvature of the bending sections was set at R = 50 mm. The distance between the waveguides in the coupling region of the multiport beam splitters was pre-calibrated and set to $d_{int}$ = 6.2 $\mu m$. Their transformation matrix was close to the recommended DFT with a fidelity of 70\% for a wavelength of 920 nm. Straight sections of length $L_{str}$ = 0.7 mm were added between the multiport beam splitters, as well as at the input and output with a length of $L_{in/out}$ = 1.143 mm. The total footprint of the waveguide circuit has sizes of 100 mm $\times$ 650 $\mu m$ = 65 mm$^2$. After writing the waveguide circuit, alignment marks (black dots) were created at the corners of the sample by ablation of the surface with an energy of 270 nJ for subsequent positioning. After that, the input and output ends of the sample were polished to optical quality. 

To implement TOPS, a titanium film with a typical thickness of 0.5 $\mu m$ was deposited onto the surface of the chip using magnetron sputtering. Next, in the same laser writing setup, the insulating tracks (blue lines) were engraved to form contact pads, supply electrodes, and heaters. The contact pads are 3.83 mm $\times$ 4 mm in size and accommodate the signal pogo-pins (red dots). Ground pogo-pins (blue dots) contact the rest of the surface. The heaters have a length of $L_{heater}$ = 6.5 mm and a width of $w_{heater}$ = 30 $\mu m$, and their resistance ranged from 1185 ohms to 1351 ohms.

\section{Thermal stability of the insertion losses}

In addition to the insertion losses measured for the zero-current configuration, insertion losses were also evaluated for two nonzero current configurations in order to assess the thermal stability of the photonic processor. Two operating conditions were considered: uniform currents of 7.8 mA and 11.7 mA applied to all heaters. These values correspond approximately to the middle and upper part of the operating range.

\begin{figure}[ht!]
\centering
\includegraphics[width=1.\linewidth]{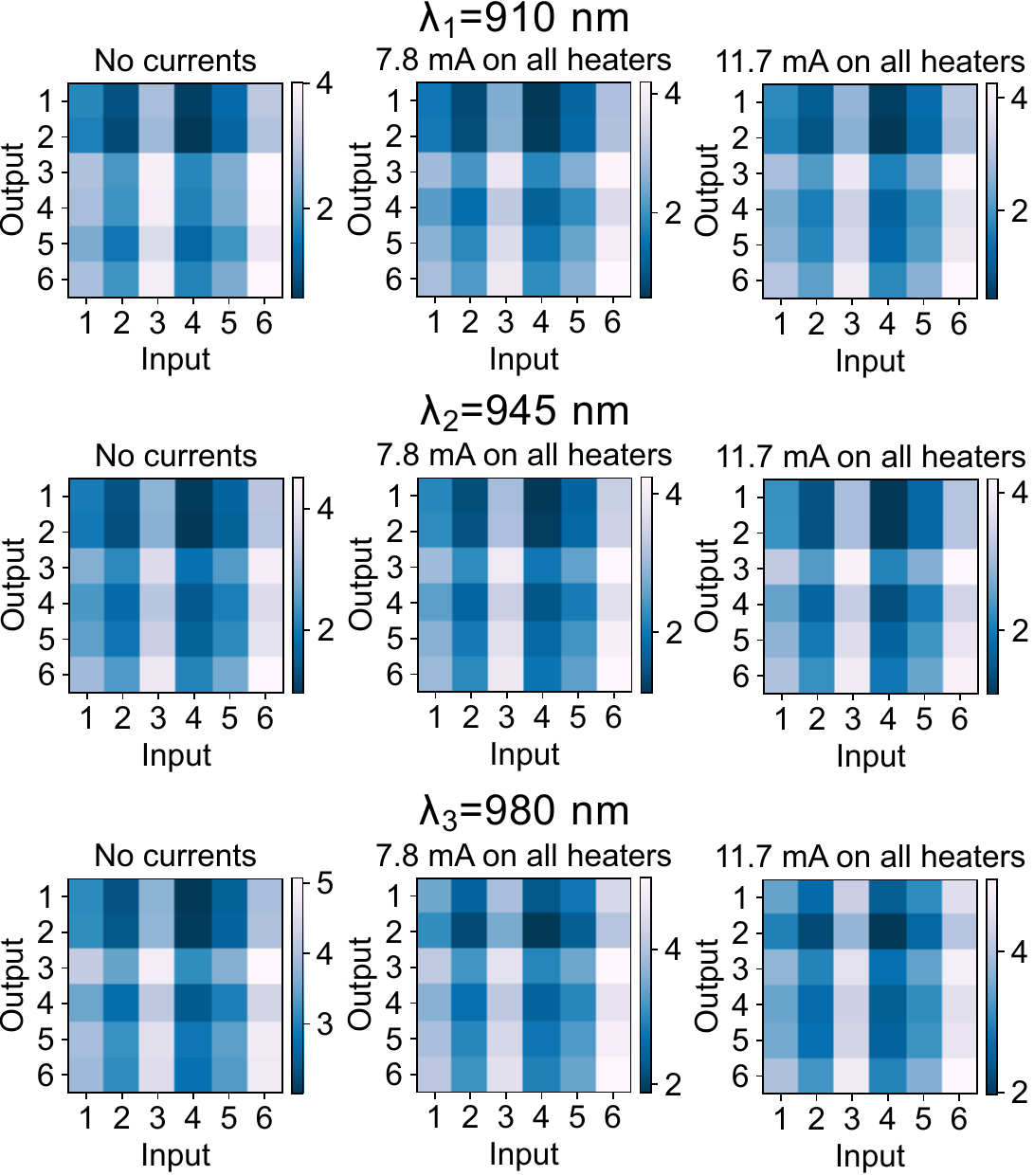}
\caption{Loss matrices measured at different wavelengths for three current configurations: zero current and uniform currents of 7.8 mA and 11.7 mA applied to all heaters. The values are given in dB.}
\label{fig:losses}
\end{figure} 

The similarity between the obtained loss matrices was quantified using the mean absolute element-wise difference:
\begin{equation}
    L(U, V) = \frac{1}{N^2}\sum_{ij}|U_{ij} - V_{ij}|.
\label{eq:loss_metric}
\end{equation}

The resulting mean differences were 0.2, 0.15, 0.1 dB for 910 nm, 0.13, 0.16, 0.1 dB for 945 nm and 0.15, 0.24, 0.15 dB for 980 nm. These low values indicate that the photonic processor remains thermally stable over the investigated range of operating currents.

\section{Estimation of crosstalk matrices} \label{app:alpha_estimation}

Estimates of the crosstalk coefficients ${\alpha_{ij}}$ were obtained by analogy with the results reported in \cite{kondratyev2025}. The resulting matrix was used both as an initial guess for the model reconstruction procedure and for assessing the physical consistency of the reconstructed parameters.

\begin{figure*}[ht!]
\centering
\includegraphics[width=1.\linewidth]{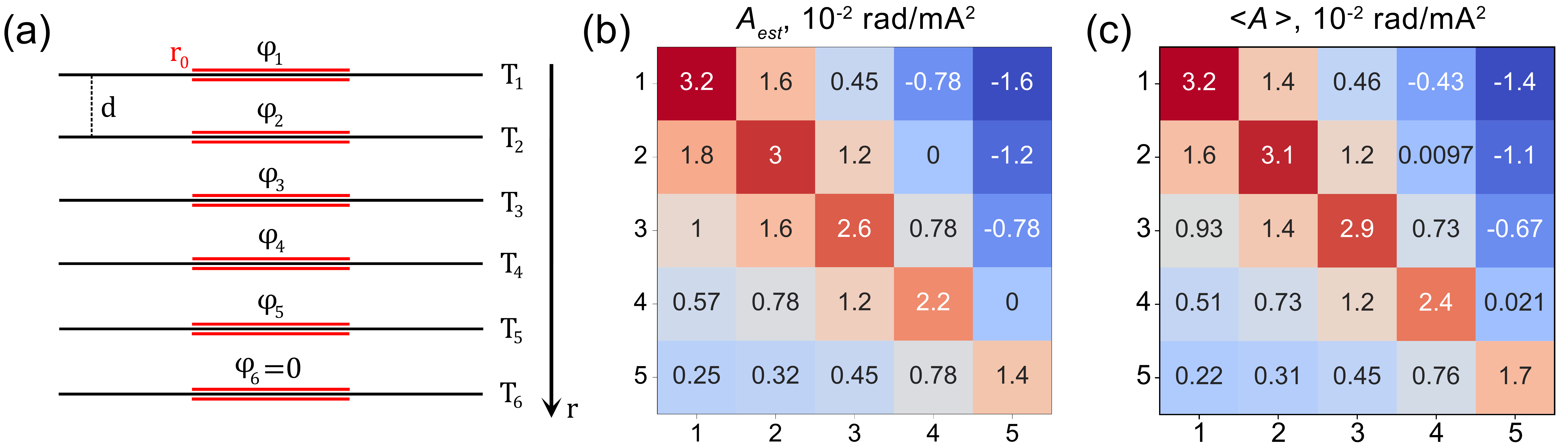}
\caption{(a) Simplified model of the phase-shifting region used to estimate the crosstalk coefficients, in which the waveguides and heaters are represented as straight structures (b) Estimated crosstalk matrix $A_{\mathrm{est}}$ (c) Averaged crosstalk matrix $\langle A\rangle$ obtained by averaging the reconstructed crosstalk matrices over all three wavelengths}
\label{fig:est_alpha}
\end{figure*}

The estimates were derived using a simplified model of the phase-shifting region (see Fig.~\ref{fig:est_alpha}a), in which the waveguides and heaters were represented as straight structures. In addition, the heater length was assumed to be much larger than its width. Under these assumptions, the heaters can be treated as line heat sources. Consequently, in a polar coordinate system $(\rho,\theta)$ centered on a heater, the source term can be represented as $q=\delta(\rho)$ \cite{flamini2015thermally}. The solution of the heat-conduction equation for the temperature distribution then takes the form

\begin{equation}\label{eq:temperature_log}
    T(r)= 
\begin{cases}
     T_0, &  |r| \leq r_0\\
     T_0 - \beta \ln \dfrac{r}{r_0}, &  |r| > r_0,
\end{cases}
\end{equation}
where $T_0$ is the heater temperature, $\beta$ is a constant with the dimension of the temperature $r_0$ is the heater width, and $r$ is the distance from the heater to the point under consideration.

Since $\phi \sim \Delta T$, it is possible to derive the dependence of the phase difference on the temperature difference for currents applied to different heaters. These relations are listed in Table~\ref{tab:phi_T}.

\renewcommand{\arraystretch}{1.2}
\begin{table*}[ht!] \centering 
\begin{tabular}{c|c|c|c|c|c} 
& 1st heater, $\alpha_{i1}$ & 2st heater, $\alpha_{i2}$ & 3st heater, $\alpha_{i3}$ & 4st heater, $\alpha_{i4}$ & 5st heater, $\alpha_{i5}$ \\ \hline 
$\phi_1$ & T(0) - T(5d) & T(d) - T(4d) & T(2d) - T(3d) & T(3d) - T(2d) & T(4d) - T(d)<0 \\ \hline 
$\phi_2$ & T(d) - T(5d) & T(0) - T(4d) & T(d) - T(3d) & T(2d) - T(2d)$\approx$0 & T(3d) - T(d)<0 \\ \hline 
$\phi_3$ & T(2d) - T(5d) & T(d) - T(4d) & T(0) - T(3d) & T(d) - T(2d) & T(2d) - T(d)<0 \\ \hline 
$\phi_4$ & T(3d) - T(5d) & T(2d) - T(4d) & T(d) - T(3d) & T(0) - T(2d) & T(d) - T(d))$\approx$0 \\ \hline 
$\phi_5$ & T(4d) - T(5d) & T(3d) - T(4d) & T(2d) - T(3d) & T(d) - T(2d) & T(0) - T(d) \\ 
\end{tabular} 
\caption{Phase-difference expressions as functions of temperature differences for currents applied to different heaters} \label{tab:phi_T} \end{table*}
\renewcommand{\arraystretch}{1.}

Recall that $\alpha_{ij}$ describes the relationship between the phase shift $\phi_i$ introduced in the $i$-th mode and the current $I$ applied to the $j$-th heater. Since $\Delta T \sim I^2$ and $\phi \sim \Delta T$, each column of Table~\ref{tab:phi_T} corresponds to a different column of the crosstalk matrix $A_m$. Therefore, the phase–temperature relations listed in Table~\ref{tab:phi_T}, together with the temperature dependence on the distance from the heater given by Eq.~\ref{eq:temperature_log}, can be used to estimate the values of $\alpha_{ij}$. As a result, the matrix $A_m$ takes the following form:

\begin{equation} \label{eq:est_A}
    A = \alpha_0\begin{pmatrix}
    \ln{\frac{5d}{r_0}} & \ln{4} & \ln{\frac{3}{2}} & -\ln{2} 
    & -\ln{4} \\
    \ln{5} & \ln{\frac{4d}{r_0}} & \ln{3} & 0 
    & -\ln{3} \\
    \ln{\frac{5}{2}} & \ln{4} & \ln{\frac{3d}{r_0}} & \ln{2} 
    & -\ln{2} \\
    \ln{\frac{5}{3}} & \ln{2} & \ln{3} & \ln{\frac{2d}{r_0}} 
    & 0 \\
    \ln{\frac{5}{4}} & \ln{\frac{4}{3}} & \ln{\frac{3}{2}} & \ln{2} 
    & \ln{\frac{d}{r_0}} \\
    \end{pmatrix},
\end{equation}
where $\alpha_0$ is a constant determined by the heater resistance, $d$ is the distance between neighboring heaters, and $r_0$ is the heater width.

The heater width is 30 $\mu$m. The maximum distance between heaters is 130 $\mu$m, while the minimum distance is approximately 80 $\mu$m. Therefore, the average distance is $d=105$ $\mu$m, corresponding to $d/r_0=3.5$. Using $\alpha_0=1.12$, obtained from the calibration of an isolated heater, yields the matrix $A_m$ shown in Fig.~\ref{fig:est_alpha}b.

The most important outcome of this estimation procedure is the relative relationship between the coefficients $\alpha_{ij}$. Their absolute values may differ from the actual ones due to uncertainties in the estimates of $\alpha_0$ and $d/r_0$, as well as the simplifying assumptions used in the model.

\section{Approximation details} \label{app:fit}

Below we provide additional details of the model reconstruction procedure.

As described in Section~\ref{subsec::models}, an experimental dataset $P(i,x_j,k)=P_{exp}(i,x_j,k)$ was measured for each wavelength. Since the measured output powers were normalized by the total output power for each input channel, the experimental data satisfy $P_{exp}(i,x_j,k)\le1,~\sum_{k}P_{exp}(i,x_j,k)=1$. Each heater was characterized using five calibration curves measured at 43 current values. Together with measurements performed for all six input channels, this yielded a dataset containing 32250 samples for each wavelength.

The interferometer model predicts the optical field amplitudes at the chip output for any current configuration. Since the measured output powers are proportional to the squared moduli of these amplitudes, the corresponding power distributions $P_{sim}((i,x_j,k))$ can be directly calculated for any current configuration. The reconstruction problem therefore consists in determining the parameter set $\{T,~M_n,~\Phi_{0m},~A_m\}$ that provides the best agreement between the predicted and experimentally measured power distributions.

Prior to optimization, the dataset was randomly shuffled and split into training and test subsets in the ratio 80:20. The model parameters were optimized on the training subset by minimizing the mean squared error (MSE):

\begin{equation}
MSE = \frac{1}{l}\sum_{ijk}(P_{exp}(i,x_j,k) - P_{sim}(i,x_j,k))^2,
\label{eq:MSE}
\end{equation}
where $l$ denotes the number of samples in the training subset. The optimization was performed simultaneously over all model parameters $\{T,~M_n,~\Phi_{0m},~A_m\}$ in Python using the PyTorch library and the Adaptive Moment Estimation (Adam) optimizer. The optimization was initialized using the estimated crosstalk matrices presented in Section~\ref{app:alpha_estimation}, discrete Fourier transform (DFT) matrices for the beam splitters $M_n$, and zero initial phases $\Phi_{0m}$.

\begin{figure*}[ht!]
\centering
\includegraphics[width=1.\linewidth]{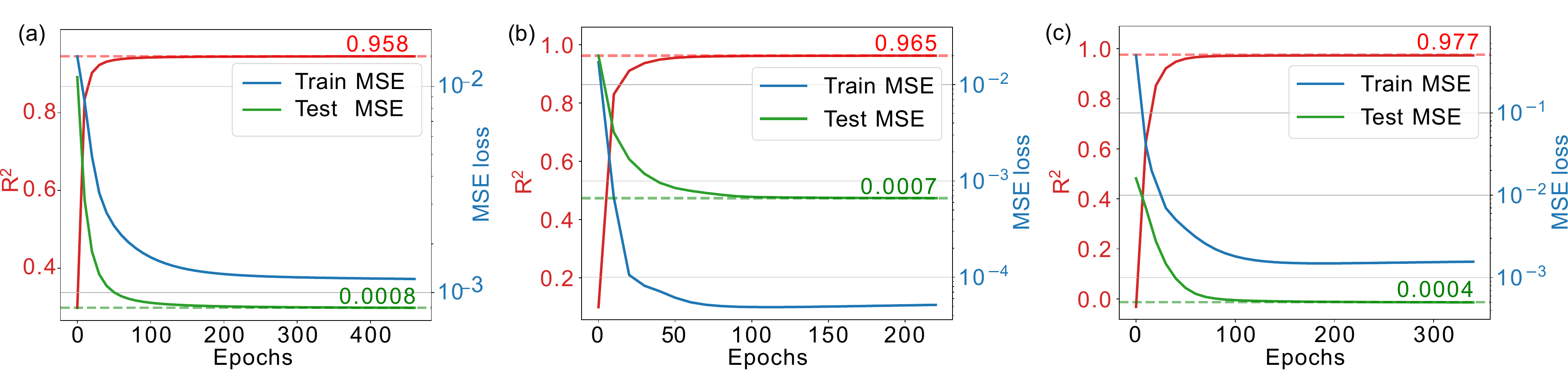}
\caption{Convergence of MSE for training and test sets and of $R^2$ for reconstructing the interferometer models at (a) 910 nm (b) 945 nm (c) 980 nm.}
\label{fig:MSE}
\end{figure*}

During optimization, the reconstruction quality was monitored using two complementary metrics: the MSE evaluated on the test subset and the coefficient of determination ($R^2$) evaluated on the training subset:

\begin{equation}
    R^2 = 1 - \frac{\sum_{ijk}(P_{exp}(i,x_j,k) - P_{sim}(i,x_j,k))^2}{\sum_{ijk}(P_{exp}(i,x_j,k) - \langle P_{exp}\rangle)^2},
\label{eq:R2}
\end{equation}
where $\langle P_{exp} \rangle$ is the average value of $P_{exp}(i,~x_j,~k)$. The coefficient of determination provides a convenient measure of the agreement between the experimental data and the model predictions. Since $R^2 = 1-MSE/Var(P_{exp})$, both metrics characterize the reconstruction quality. During optimization, the MSE on the test subset and the value of $R^2$ on the training subset were monitored simultaneously. The optimization was terminated once the increase in the coefficient of determination between successive epochs fell below $10^{-5}$. The corresponding convergence curves are shown in Fig.~\ref{fig:MSE}. 

In all cases, the optimization converged reliably, reaching stable values of both the MSE and the coefficient of determination.

\section{Simultaneous heater calibration}\label{app:simultaneous_calibration_5X}

In the conventional calibration measurement procedure described in Section \ref{subsec::calib}, for each input port of the optical chip (of 6), 25 separate calibration sweeps are measured: all 25 active heaters are swept one by one individually. This results in a total of 150 experimentally measured current sweeps, which form the total experimental dataset for the digital model to be trained on. In this scenario, given that an individual current sweep for each heater takes approximately 50 seconds (43 current steps with a 1-second duration each and a 7-second time break for thermal cooling after), the measurement of the entire calibration data set takes slightly more than two hours and 13 minutes. This time could be significantly reduced by exploiting a simultaneous calibration measurement strategy in which several heaters are swept at once. In our case, five heaters are swept simultaneously, one from each phase layer. This procedure is repeated five times (for each input port of the chip), with a different set of sweeping heaters each time. Therefore, there are a total of 5 simultaneous current sweeps or calibration measurements for each of the input ports of the chip. This results in a total of 30 experimentally measured current sweeps instead of 150 for the conventional procedure. We would like to emphasize that both of these data sets contain an equal amount of information about the physical model of the optical chip. The simultaneous sweep strategy is simply a denser container. For this reason, we will refer to the simultaneous calibration strategy as dense calibration in the following.

Speaking of the time profit, for dense calibration, it can be up to five times faster than the one by one calibration in the limit. However, we chose a slightly less extreme option. First, we have increased the number of current sweeps to 51 due to the increased information density in the calibration data. Second, we have set the time duration of a single sweep to 2 seconds. After changing the current on each of the five active heaters, we wait 2 seconds for the system to thermally relax. With these modifications, the entire dense calibration process takes approximately 55 minutes, instead of 2 hours and 13 minutes, which is still a significant improvement in speed.

Moreover, despite the explicit acceleration of the measurement time, dense calibration offers an additional advantage over the conventional calibration procedure. This is because the experimental data obtained is five times denser, and it consequently takes five times less time to process, which leads to faster post-processing and model parameter retrieval. 

\begin{figure*}[ht!]
\centering
\includegraphics[width=1.\linewidth]{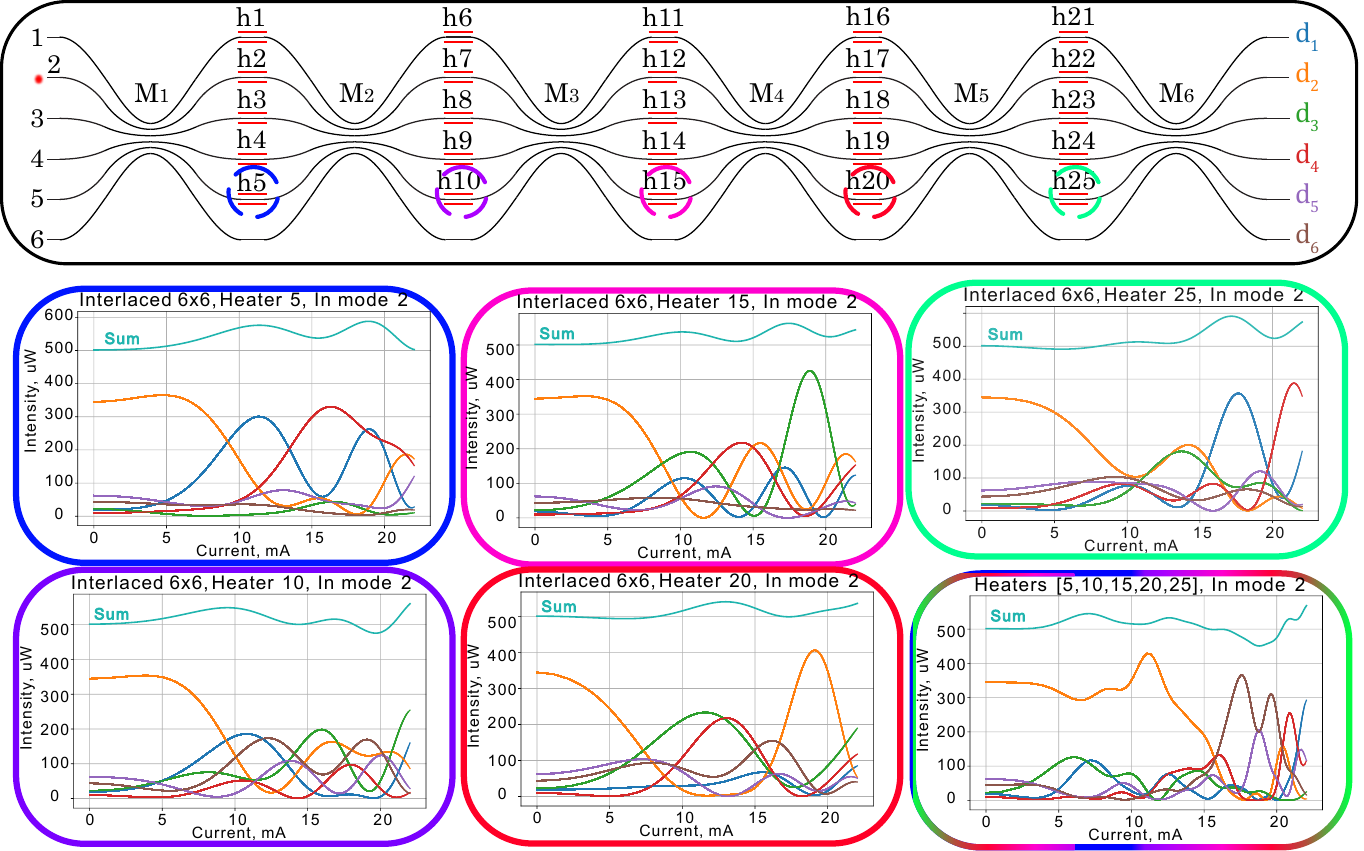}
\caption{Simultaneous calibration measurements illustration. Instead of 5 separately measured current sweeps of heaters 5, 10, 15, 20, and 25, one can simultaneously sweep these heaters all together while registering the output optical power signal.}
\label{fig:dense_calib}
\end{figure*}

Figure \ref{fig:dense_calib} illustrates a dense calibration routine: instead of measuring 5 separate current sweeps of heaters 5, 10, 15, 20, and 25, we simultaneously sweep these heaters all together while registering the optical power output signal. Another example of a measured sweep of five heaters simultaneously is shown in Fig.\ref{fig:R_2_Distributions} (d, e). All experimental calibration data with corresponding digital models are available at \cite{NN_Interferometer}.

\textcolor{black}{\section{Determination of current configurations for target unitary transformations} \label{app:found}}
\textcolor{black}{The current configuration $\vec{x}$ required to implement a target unitary transformation $U_{target}$ was determined using the reconstructed model in two consecutive stages. First, we determined the internal phase shifts for which the interferometer implements $U_{target}$ up to arbitrary input and output phase shifts. The resulting phases were then converted into physically realizable heater currents using the calibrated phase-current relation. Thus, the first stage was performed entirely in the phase domain, while the second stage provided the corresponding physical control parameters.}

\textcolor{black}{For the first stage, we introduce auxiliary input and output phase layers and write the transformation used in the optimization as
\begin{equation}
    U_{theory} = P_{out}\cdot U\cdot P_{in} = P_{out}(M_6 \prod_{i=1}^{5} (P_i M_i)) P_{in}
\end{equation},
where $P_{in/out} = diag(e^{i\theta^{in/out}_{1}}, \dots,~e^{i\theta^{in/out}_{5}}, 1)$, $\theta^{in/out}_{j}, j=\overline{1,5}$ - arbitrary input and output phase shifts. These matrices correspond to the input and output phase layers of the universal interferometer architecture \cite{Robust2020}. Such layers were not physically implemented in the fabricated chip and are introduced here only as auxiliary degrees of freedom accounting for the ambiguity of the input and output phases. These phases do not affect the measured output power distributions.}

\textcolor{black}{The optimization therefore involved 35 phase parameters: 25 internal phase shifts of the fabricated interferometer and 10 auxiliary input and output phase shifts. They were determined by minimizing the infidelity Eq.~\ref{eq:fid}:
\begin{equation}
    \mathcal{L}_1=1 - F_1 = 1-\frac{1}{N}|Tr(U_{target}^{\dagger}U_{theory})|.
\end{equation},
where $N=6$. The optimization was performed using the basinhopping algorithm from scipy.optimize. After convergence, the auxiliary input and output phases were discarded, leaving the 25 internal phase shifts.}

\textcolor{black}{The second stage converts these internal phase shifts into the corresponding heater currents. An important point is that the first-stage optimization depends on the phases only through the factors $e^{i\phi}$. Consequently, each obtained phase is defined modulo $2\pi$ : replacing any phase $\phi$ by $\phi+2\pi k$ where $k$ is an integer, leaves the implemented optical transformation unchanged. This freedom can be used to select a phase representation that corresponds to physically realizable heater currents.}

\textcolor{black}{The 25 internal phase shifts were divided into five vectors $\Phi_m, m=\overline{1,5}$, corresponding to the five phase layers. For each layer,  Eq.~\ref{eq:phases_mtx} can be written as
\begin{equation}
\Phi_m = \Phi_{0m} + A_m\cdot X_m^2,
\label{eq:phases_layer}
\end{equation}}

\begin{figure*}[ht!]
\centering
\includegraphics[width=1.\linewidth]{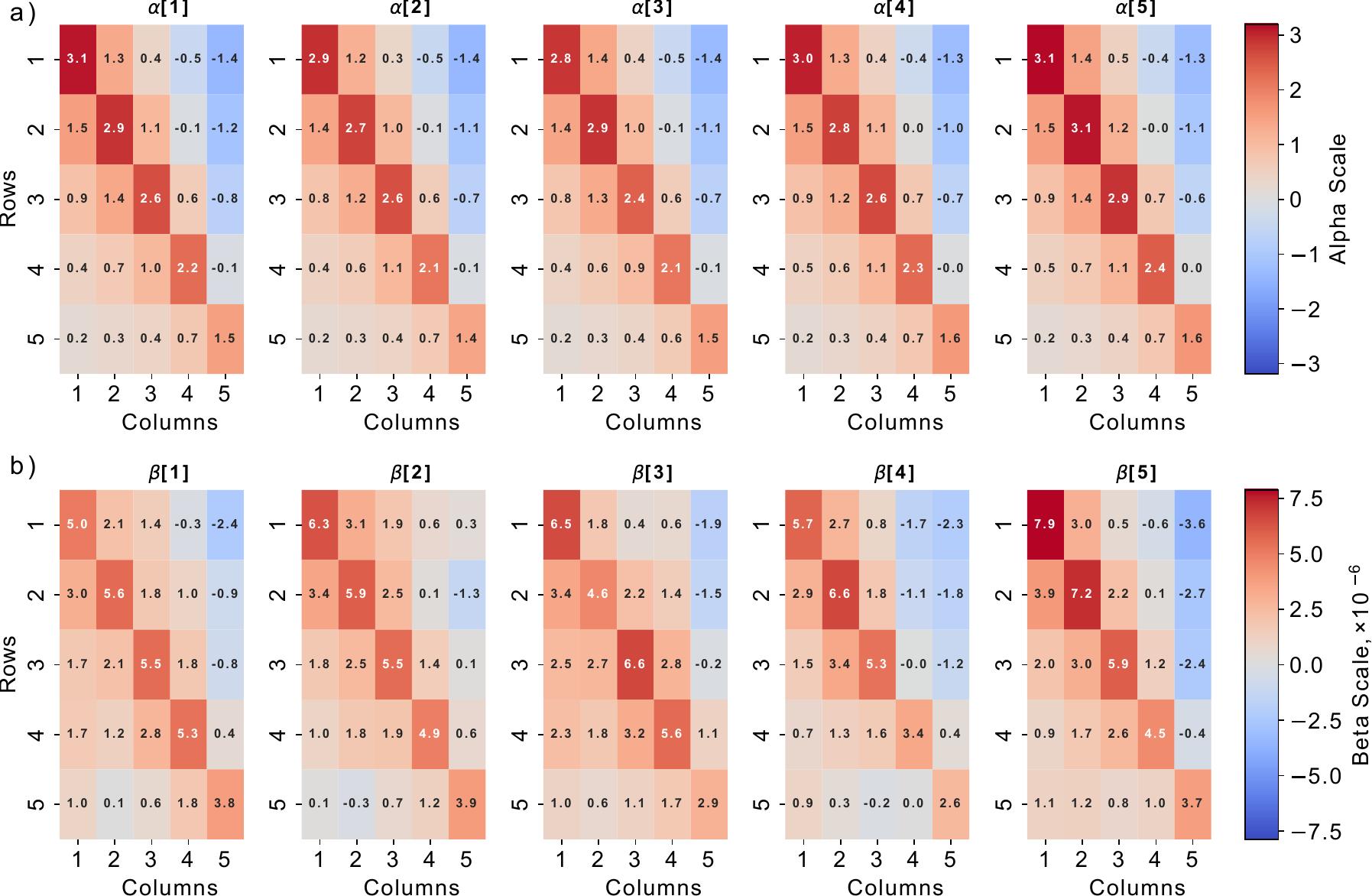}
\caption{ The parameter values, $A$ and $B$, obtained for the optical chip model (\ref{eq:phases_mtx_fourth}) at a wavelength of 925 nm. These parameters correspond to the $R^2$ distribution shown in Fig. \ref{fig:R_2_Distributions}(a). }
\label{fig:all_params_forth_25}
\end{figure*}

\textcolor{black}{where $X=(x_1,\dots,x_5)^T$ is the current vector for the corresponding phase layer. Introducing the squared-current vector $Y_m=X^2_m$ gives
\begin{equation}
    Y = A^{-1}(\Phi - \Phi_0).
\end{equation}}
\textcolor{black}{Direct evaluation of this expression may result in negative components of $Y_m$, which cannot correspond to real heater currents. To exploit the $2\pi$-periodicity of the phases, we therefore introduce an integer-valued vector $K_m=(k_1,\dots,k_5)^T$ whose components specify the number of $2\pi$ periods added to the corresponding target phases. The squared-current vector can then be written as
\begin{equation}
    Y = A^{-1}(\Phi - \Phi_0 + 2\pi K_m).
    \label{eq:phx_to_currs}
\end{equation}}

\textcolor{black}{Since all components of $Y_m$ must be nonnegative, an appropriate $K_m$ was determined iteratively. The procedure was initialized with $K_m=(0,\dots,0)$. If the resulting $Y_m$ contained negative components, the component of $K_m$ corresponding to the first negative element of $Y_m$ was increased by one, thereby adding $2\pi$ to the corresponding phase. The squared-current vector was then recalculated using Eq.~\ref{eq:phx_to_currs}. This procedure was repeated until all components of $Y_m$ became nonnegative.}

\textcolor{black}{The corresponding heater currents were finally obtained element-wise as 
\begin{equation}
    X_m = \sqrt{Y_m}.
\end{equation}}

\textcolor{black}{Repeating this procedure independently for all five phase layers and concatenating the resulting vectors $X_m$ yielded the complete 25-element current vector $\vec{x}$ required to implement $U_{target}$.}

\section{Numerical optimization of multiwavelength routing}
\label{app:rout}

The optimization procedure used to determine the current configurations for multiwavelength routing is described below. We first formulate the routing problem in a general form, which includes both multiplexing and demultiplexing as particular cases.

Let radiation at wavelengths $\lambda_1$, $\lambda_2$, and $\lambda_3$ be injected into input channels $i_1$, $i_2$, and $i_3$, respectively, and routed to output channels $j_1$, $j_2$, and $j_3$. For a given wavelength $\lambda_k$, radiation injected into the $i_k$-th input channel is transformed according to the $i_k$-th column of the interferometer matrix $U_{\lambda_k}$. Therefore, the general three-wavelength routing problem can be formulated as minimization of

\begin{equation}
\mathcal{L}_2(\vec{x})=\sum_{k=1}^{3}\left\|\left|U_{\lambda_k}^{\,i_k}(\vec{x})\right|^2-O^{j_k}\right\|_2,
\label{eq:routing_general}
\end{equation}

where $U_{\lambda_k}^{\,i_k}$ denotes the $i_k$-th column of the transformation matrix at wavelength $\lambda_k$, $\vec{x}$ is the current configuration, and $O^{j_k}$ is a column vector with a 1 at the $j_k$-th position and zeros elsewhere. Thus, the optimization constrains only the output power distributions corresponding to the selected input channels, while the remaining columns of the transformation matrices may take arbitrary values.

Demultiplexing and multiplexing are particular cases of Eq.~\ref{eq:routing_general}. For demultiplexing, radiation at all three wavelengths is injected into the same input channel, $i_1=i_2=i_3$, and routed to different output channels. For multiplexing, radiation at the three wavelengths is injected into different input channels, while the target output channel is the same, $j_1=j_2=j_3$.

The optimization was performed using multistart L-BFGS-B initialized with a scrambled Sobol sequence, followed by discrete integer refinement. The optimization variables were the 25 currents applied to the phase shifters. As follows from Eq.~(4), the phase shifts depend quadratically on the applied currents. Therefore, instead of optimizing directly over the currents $x_n$, we introduced squared-current variables

\begin{equation}
p_n\propto x_n^2,\qquad n=1,\ldots,25.
\label{eq:squared_current}
\end{equation}

In terms of the squared-current variables, Eq.~(4) becomes linear in the optimization variables for each phase layer. This transformation converts the quadratic current dependence of the phase into a linear dependence on the new optimization variables.

Since the routing functional has many local minima, the result of a single L-BFGS-B optimization can depend strongly on the initial point. To provide broad coverage of the 25-dimensional parameter space, 2048 initial points were therefore generated using a scrambled Sobol low-discrepancy sequence. L-BFGS-B optimization was performed from each initial point using the analytical gradient of Eq.~\ref{eq:routing_general}. The gradient was calculated directly from the interferometer model by differentiating the propagation through the phase and mixing layers with respect to the squared-current variables.

To reduce the computational cost of the multistart procedure, successive screening of the initial points was used. The 2048 candidates were first subjected to a short L-BFGS-B optimization with a maximum of 5 iterations, after which the 1024 candidates with the lowest values of $\mathcal{L}$ were retained. Two further screening stages with maximum iteration counts of 8 and 12 reduced the number of candidates to 384 and 128, respectively. The 64 best candidates remaining after the screening procedure were then optimized with L-BFGS-B until convergence.

L-BFGS-B operates in the continuous squared-current space, whereas the current source requires discrete current settings. The optimized values of $\vec{p}$ were therefore converted back to current values and rounded to the nearest allowed integer settings. Since rounding can slightly increase the routing functional, the resulting current configurations were additionally refined directly on the discrete grid. The best candidates were first tested using the neighboring integer values corresponding to rounding in either direction, and were subsequently subjected to a coordinate-wise search with decreasing current steps of 5, 3, 2, and 1. The refinement was terminated when no further decrease of the original routing functional could be obtained.

The complete procedure therefore combines a broad sampling of the parameter space with gradient-based local optimization and a final search over experimentally accessible current values. Although this procedure does not guarantee identification of the global minimum, the use of multiple broadly distributed initial points substantially reduces the dependence of the obtained solution on a particular initial current configuration.

\section{Further physical model improvements}
\label{app:fouth}

An example of a set of experimental calibration data at a wavelength of 925 nanometers is presented in Fig. \ref{fig:R_2_Distributions}(d-e). The left panel shows the fit of the experimental data with a model that has a quadratic dependence of the phase on the current, while the right panel shows the fit with a model having a fourth-degree dependence of the phase on current. The measured intensities, normalized by the sum of the output from the optical chip, are represented by the currents, and the values predicted by the models are represented by the continuous lines. The legend displays the values of the coefficient of determination $R^2_j$ for each experimental curve. Figures \ref{fig:R_2_Distributions} (a-c) show the histograms of the distribution of these coefficients. This example demonstrates an improvement in the quality of the data for the model with the fourth-degree phase-current dependence compared to the model with only a quadratic dependence.

The parameter values obtained for the improved optical chip model at a wavelength of 925 nm are shown in Fig. \ref{fig:all_params_forth_25}. All experimental parameters for each model are available at \cite{NN_Interferometer}.

\end{document}